\RequirePackage{fix-cm}
\documentclass[twocolumn,sort,compress]{svjour3}          

\smartqed 

\usepackage{epsfig} 
\usepackage{xcolor}
\definecolor{azure}{rgb}{0.0, 0.5, 1.0}
\definecolor{awesome}{rgb}{1.0, 0.13, 0.32}
\definecolor{asparagus}{rgb}{0.53, 0.66, 0.42}
\definecolor{cadetgrey}{rgb}{0.57, 0.64, 0.69}

\usepackage{amsmath,amssymb}
\usepackage{xparse}
\usepackage{tikz}
\usetikzlibrary{trees}
\usetikzlibrary{mindmap}
\usetikzlibrary{matrix,backgrounds}
\pgfdeclarelayer{myback}
\pgfsetlayers{myback,background,main}
\tikzset{mycolor/.style = {line width=1bp,color=#1}}
\tikzset{myfillcolor/.style = {draw,fill=#1,#1,rounded corners}}

\NewDocumentCommand{\fhighlight}{O{blue!40} m m}{%
\draw[myfillcolor=#1] (#2.north west)rectangle (#3.south east);}
\NewDocumentCommand{\fhighlightL}{O{blue!40} m m}{%
\draw[myfillcolor=#1] (#2.south west)rectangle (#3.south east);}

\usepackage{xfrac}
\usepackage{url}
\usepackage{subcaption}
\usepackage{siunitx}
\usepackage{booktabs}
\usepackage[hidelinks]{hyperref}
\usepackage[sort&compress,numbers]{natbib}
\usepackage{bm}
\usepackage{enumitem}
\usepackage{nomencl}
\makenomenclature

\def\intdomd{\Omega_\delta}
\def\uelast{\underline{u}}
\def\uperid{u}

\graphicspath{{./Figures/}}
\usepackage{etoolbox}

\usepackage[british]{babel}

\begin{document}

\title{Coupling approaches with non-matching grids for classical linear elasticity and bond-based peridynamic models in 1D}
\titlerunning{Coupling approaches with non-matching grids}

\author{Patrick Diehl \and Emily Downing \and Autumn Edwards \and Serge Prudhomme}

\institute{P. Diehl \at
Applied Computer Science (CCS-7) \\
Los Alamos National Laboratory\\
Center of Computation \& Technology\\
and \\
Department of Physics and Astronomy\\
Louisiana State University\\
Baton Rouge, Louisiana 70803\\
\email{diehlpk@lanl.gov} \\
Url: \url{https://orcid.org/0000-0003-3922-8419}
\and 
E. Downing \at
Massachusetts College of Liberal Arts\\
375 Church Street
North Adams, MA 01247, USA\\
\email{ed0572@mcla.edu}
\and
A. Edwards \at
Howard University \\
2400 6th St NW, Washington, DC 20059, USA\\
\email{autumn.edwards@bison.howard.edu}
\and
S. Prudhomme \at
Department of Mathematics and Industrial Engineering\\ 
Polytechnique Montr\'eal, \\
C.P. 6079, succ.\ Centre-ville, Montr\'eal, Qu\'ebec H3C~3A7, Canada \\
\email{serge.prudhomme@polymtl.ca} \\
Url: \url{ https://orcid.org/0000-0002-1168-5290}
}

\date{Received: date / Accepted: date}
\maketitle    
\begin{abstract}
Local-nonlocal coupling approaches provide a means to combine the computational efficiency of
local models and the accuracy of nonlocal models. To facilitate the coupling of the two models, non-matching grids are often desirable as nonlocal grids usually require a finer resolution than local grids. In that case, it is often convenient to resort to interpolation operators so that models can exchange information in the overlap regions when nodes from the two grids do not coincide. This paper studies three existing coupling approaches, namely 1) a method that enforces matching displacements in an overlap region, 2) a variant that enforces a constraint on the stresses instead, and 3) a method that considers a variable horizon in the vicinity of the interfaces. The effect of the interpolation order and of the grid ratio on the performance of the three coupling methods with non-matching grids is carefully studied on one-dimensional examples using polynomial manufactured solutions. The numerical results show that the degree of the interpolants should be chosen with care to avoid introducing additional modeling errors, or simply minimize these errors, in the coupling approach. 

\end{abstract}

\section{Introduction}
\label{sec:introduction}

In the past few years, there has been a great interest in the development of local-nonlocal coupling methods for computer simulations of problems in solid mechanics, see \emph{e.g.}\ the recent survey~\cite{d2021review} and references therein. There are two main advantages in using a coupling approach. First, it can be employed to apply the boundary conditions associated with a nonlocal model. Indeed, one challenge in nonlocal modeling, as highlighted in the survey~\cite{diehl2019review} on validation experiments for peridynamics models, lies in the treatment of the nonlocal boundary conditions, see \emph{e.g.}\textcolor{blue}{~\cite{du2016nonlocal, madenci2018state, madenci2018weak, gu2018revisit, Prudhomme-Diehl-2020, you2020asymptotically, d2021prescription, d2020physically,zhao2025enforcing}}. Using a coupling approach, boundary conditions can be prescribed in terms of the local model so that the nonlocal model is considered only in the interior of the domain. Second, it can help reduce computational costs, since non-local models, such as peridynamics, molecular dynamics, or smoothed particle hydrodynamics, can be very computationally intensive, see for instance~\cite{diehl2022comparative}. 

A vast literature on coupling methods for peridynamics and linear elasticity is now available, see \emph{e.g.}~\cite{d2021review}. However, the large majority of the discrete approaches involving an overlapping region consider the grid spacing in the local and nonlocal models to be the same and the grid points to coincide, apart from a few contributions~\cite{zaccariotto2018coupling,zaccariotto2017enhanced}. Yet, the scales introduced by the two models may be quite different, so that it is desirable to consider a different nodal spacing, using large grid sizes for the local model and smaller ones for the nonlocal model. Moreover, for the sake of flexibility in the design of the two grids, it would also be desirable that the grid points at the interface of the two regions do not necessarily coincide. For non-matching grid configurations, it is then necessary to resort to interpolation or projection operators of the displacement fields so that information from the two models can be properly exchanged at their interface. More importantly, it is essential that the order of accuracy of the interpolation or projection operators be sufficiently high to retain the approximation order of the discrete methods used to obtain the numerical solutions to the local and nonlocal models in their respective regions. The main contributions of this paper are the development of interpolation operators for coupling linear elasticity and peridynamics models in the case of non-matching grids and an extensive study of their influence on the accuracy of the coupled solutions. 

Various coupling approaches have been proposed in the past, based either on matching the displacements or stresses obtained from the two models over an overlapping region, or on shrinking the size of the nonlocal horizon when transitioning from the nonlocal to the local model, at which point the horizon vanishes. In this paper, we will consider the coupling methods described in the earlier paper~\cite{diehl2022coupling}, namely the matching displacement method (MDCM), the matching stress coupling method (MSCM), and the variable horizon coupling method (VHCM). For the sake of simplicity in the exposition, we will consider a model problem based on a one-dimensional bar with constant and homogeneous material properties. However, numerical results will be extended to the case of a material with varying stiffness in order to mimic potential defects that would require the use of a nonlocal model such as peridynamics.

The paper is organized as followed: Section~\ref{sec:model:preliminaries} descri\-bes the formulations of the classical linear elasticity model and the bond-based peridynamic model for the one-dimensional bar problem, as well as preliminary notions and notations that will be useful in the remainder of the paper. Section~\ref{sec:coupling:theory} introduces the three coupling methods for linear elasticity and peridynamics in the continuous setting. The presentation will essentially summarizes that of~\cite{diehl2022coupling}. In Section~\ref{sec:coupling:discrete}, the discrete counterparts of the coupling methods with non-matching grids are introduced. In particular, the interpolation operators will be described in detail. We present a series of numerical examples in  Section~\ref{sec:numericalexamples} before providing some concluding remarks in Section~\ref{sec:conclusion}.    

\section{Model problem and preliminaries}
\label{sec:model:preliminaries}

The model problem deals with the quasi-static simulation of a one-dimensional bar subjected to mixed boundary conditions (Dirichlet and Neumann) at the extremities of the bar. We assume that the deformations are small enough to be described by linear elasticity. We introduce in this section the local model based on classical elasticity theory and the nonlocal model based on the linearized bond-based peridynamic theory~\cite{Silling-JMPS-2000}.

\subsection{Classical linear elasticity model}

Let $\ell$ denote the length of the bar. We thus consider the domain $\Omega = (0,\ell) \subset \mathbb{R}$ and write $\bar{\Omega}=[0,\ell]$. The local problem consists in finding the displacement $\underline{u} = \underline{u}(x)$, $\forall x \in \bar{\Omega}$, such that:
\begin{align}
\label{eq:1dlinearelasticity}
- \big( EA \uelast' \big)' = f_b, &\quad \forall x \in \Omega, \\
\label{eq:Dirichlet}
\uelast = 0, &\quad \text{at}\ x=0,\\
\label{eq:Neumann}
EA\uelast' = g, &\quad \text{at}\ x=\ell,
\end{align}
where $E=E(x)$ is the modulus of elasticity, $A=A(x)$ denotes the cross-sectional area of the bar, $f_b=f_b(x)$ is the external body force density (per unit length), and $g\in \mathbb{R}$ is a traction force applied at $x=\ell$. We assume that $f_b$ is chosen smooth enough so that the solution $u$ can be differentiated as many times as needed. 

For the sake of simplicity in the presentation, we shall consider that $E$ and $A$ are constant along the bar and take $A$ equal to unity (if $A$ is different from unity, it suffices to replace all occurrences of $E$ by $EA$). In that case,  Equation~\eqref{eq:1dlinearelasticity} can be simplified as
\begin{equation}
- E \underline{u}'' = f_b, \quad \forall x \in \Omega .
\end{equation}
Moreover, we consider here a Dirichlet boundary condition at one end and a Neumann boundary condition at the other hand -- Problem~\eqref{eq:1dlinearelasticity}-\eqref{eq:Neumann} is thus referred to as a mixed boundary-value problem -- to keep the presentation as general as possible. However, in the numerical results, we will also deal with problems with homogeneous Dirichlet boundary conditions at both ends and thus replace~\eqref{eq:Neumann} by $\underline{u}(\ell)=0$. 

\subsection{Linearized bond-based peridynamic model}

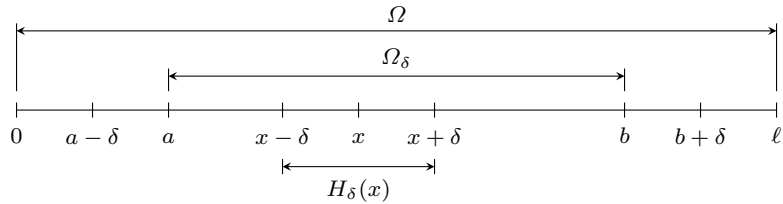
\begin{figure*}[tbp]
\centering
\small
\begin{tikzpicture}
\draw (0,0) -- (10.0,0);
\draw (0,-0.1) -- (0,0.1);
\draw (1,-0.1) -- (1,0.1);
\draw (2,-0.1) -- (2,0.1);
\draw (3.5,-0.1) -- (3.5,0.1);
\draw (4.5,-0.1) -- (4.5,0.1);
\draw (5.5,-0.1) -- (5.5,0.1);
\draw (8.0,-0.1) -- (8.0,0.1);
\draw (9.0,-0.1) -- (9.0,0.1);
\draw (10.0,-0.1) -- (10.0,0.1);
\node[above] at (0.0,-0.55) {$0$};
\node[above] at (1.0,-0.59) {$a-\delta$};
\node[above] at (2.0,-0.55) {$a$};
\node[above] at (3.5,-0.59) {$x-\delta$};
\node[above] at (4.5,-0.55) {$x$};
\node[above] at (5.5,-0.59) {$x+\delta$};
\node[above] at (8.0,-0.55) {$b$};
\node[above] at (9.0,-0.59) {$b+\delta$};
\node[above] at (10.0,-0.55) {$\ell$};
\draw[arrows=<->, >=stealth]  (0.0,1.05) -- (10.0,1.05);
\draw  (0.0,0.25) -- (0.0,1.15);
\draw  (10.0,0.25) -- (10.0,1.15);
\node[above] at (5.00,1.05) {$\Omega$};
\draw[arrows=<->, >=stealth]  (2.0,0.45) -- (8.0,0.45);
\draw  (2.0,0.25) -- (2.0,0.55);
\draw  (8.0,0.25) -- (8.0,0.55);
\node[above] at (5.00,0.45) {$\intdomd$};
\draw[arrows=<->, >=stealth]  (3.5,-0.75) -- (5.5,-0.75);
\draw  (3.5,-0.65) -- (3.5,-0.85);
\draw  (5.5,-0.65) -- (5.5,-0.85);
\node[below] at (4.5,-0.8) {$H_\delta(x)$};
\end{tikzpicture}
\caption{Definition of the computational domains for the peridynamic model (adapted from~\cite{diehl2022coupling}).}
\label{Fig:peridynamicsdomains}
\end{figure*}

Let $\Omega_\delta=(a,b)$ be the subregion of domain $\Omega$, where one would like to describe the material using a nonlocal model, see \emph{e.g.}\ Figure~\ref{Fig:peridynamicsdomains}. The peridynamic steady-state equation of the linearized micro-elas\-tic model~\cite{Silling-JMPS-2000} reads for any dimension $d=1$, $2$, or $3$,
\begin{equation}
\label{eq:linearizedperidynamics}
- \int_{H_\delta(x)} \kappa(x) \frac{\xi \otimes \xi}{\| \xi \|^3} \big(\uperid(y) - \uperid(x)\big) dy = f_b(x),
\end{equation}
where $\delta > 0$ is the so-called horizon, $H_\delta(x)$ is the neighborhood of point $x$, that is,
\[
H_\delta(x) = \{y\in \mathbb R^d;\ \| y-x\| < \delta \},
\]
$\kappa=\kappa(x)$ is the peridynamic stiffness parameter bet\-ween point~$x$ and a neighboring point $y\in H_\delta(x)$, $\xi=y-x$ is the vector between the two points $x$ and $y$ in the reference configuration, $\Vert \cdot \Vert$ is the Euclidean norm, and $u$ is the displacement in the deformed configuration. We note that the displacement solution for the peridynamic model is denoted by $u$ to distinguish it from the displacement~$\underline{u}$ obtained by the classical linear elasticity model. For the one-dimensional case, the above equation can be rewritten as
\begin{equation}
\label{eq:1dperidynamics}
- \int_{x-\delta}^{x+\delta} \kappa(x) \frac{\uperid(y) - \uperid(x)}{|y-x|} dy = f_b(x).
\end{equation}
where in this case, the neighborhood of $x$ is given as the open interval $H_\delta(x)=(x-\delta,x+\delta)$. One approach to identify the peridynamic stiffness constant $\kappa$ is by matching the strain energy in both models~\cite{Silling-JMPS-2000}, so that 
\begin{equation}
\frac{\kappa(x)\delta^2}{2} = E(x), 
\end{equation}
or, equivalently,
\begin{equation}
\label{eq:kappa:variable}
\kappa(x) = \frac{2E(x)}{\delta^2}.
\end{equation}
If the Young's modulus $E$ is assumed constant over the whole domain, the peridynamic stiffness constant simply reads
\begin{equation}
\label{eq:kappa}
\kappa = \frac{2E}{\delta^2}.
\end{equation}
Finally, to avoid any issue with the integral in~\eqref{eq:1dperidynamics} at the boundary of $\Omega$, we assume that $\Omega_\delta=(a,b)$ is such that $a>\delta$ and $b<\ell-\delta$.

\section{Coupling methods}
\label{sec:coupling:theory}

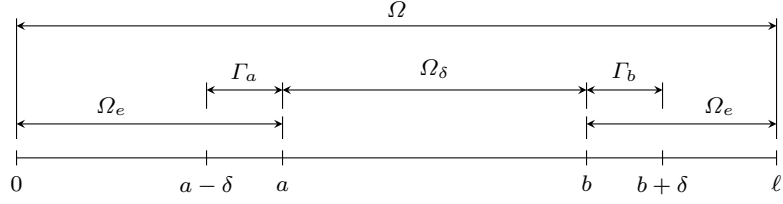
\begin{figure*}[tbp]
\centering
\begin{tikzpicture}
\draw (0,0) -- (10.0,0);
\draw (0,-0.1) -- (0,0.1);
\draw (2.5,-0.1) -- (2.5,0.1);
\draw (3.5,-0.1) -- (3.5,0.1);
\draw (7.5,-0.1) -- (7.5,0.1);
\draw (8.5,-0.1) -- (8.5,0.1);
\draw (10.0,-0.1) -- (10.0,0.1);
\node[above] at (0.0,-0.55) {$0$};
\node[above] at (2.5,-0.59) {$a-\delta$};
\node[above] at (3.5,-0.55) {$a$};
\node[above] at (10.0,-0.55) {$\ell$};
\node[above] at (7.5,-0.55) {$b$};
\node[above] at (8.5,-0.59) {$b+\delta$};
\draw[arrows=<->, >=stealth]  (0.0,1.75) -- (10.0,1.75);
\draw  (0.0,0.25) -- (0.0,1.85);
\draw  (10.0,0.25) -- (10.0,1.85);
\node[above] at (5.00,1.75) {$\Omega$};
\draw[arrows=<->, >=stealth]  (0.0,0.45) -- (3.5,0.45);
\draw  (3.5,0.25) -- (3.5,0.55);
\node[above] at (1.25,0.45) {$\Omega_e$};
\draw[arrows=<->, >=stealth]  (2.5,0.9) -- (3.5,0.9);
\draw  (2.5,0.65) -- (2.5,1.0);
\node[above] at (3.0,0.9) {$\Gamma_a$};
\draw[arrows=<->, >=stealth]  (3.5,0.9) -- (7.5,0.9);
\draw  (3.5,0.65) -- (3.5,1.0);
\node[above] at (5.5,0.94) {$\intdomd$};
\draw  (7.5,0.65) -- (7.5,1.0);
\draw[arrows=<->, >=stealth]  (7.5,0.9) -- (8.5,0.9);
\draw  (8.5,0.65) -- (8.5,1.0);
\node[above] at (8.0,0.9) {$\Gamma_b$};
\draw[arrows=<->, >=stealth]  (7.5,0.45) -- (10.0,0.45);
\draw  (7.5,0.25) -- (7.5,0.55);
\node[above] at (9.25,0.45) {$\Omega_e$};
\end{tikzpicture}
\caption{Definition of the computational domains for coupling the linear elasticity model with the bond-based peridynamic model (adapted from~\cite{diehl2022coupling}).}
\label{Fig:couplingLinearElastPeryd}
\end{figure*}

We describe in this section the three coupling methods previously presented in~\cite{diehl2022coupling} for the configuration shown in Figure~\ref{Fig:couplingLinearElastPeryd}. We consider that the peridynamic model is chosen in the region $\Omega_\delta = (a,b)$ while the classical linear elasticity model is employed in $\Omega_e = (0,a) \cup (b,\ell)$. Depending on the coupling method, the two models may overlap in the two regions $\Gamma_a = (a-\delta,a)$ and $\Gamma_b = (b,b+\delta)$. Moreover, we will suppose in this section that $E$ and $\kappa$ are constant over the whole domain.

\subsection{Matching-displacement coupling method (MDCM)}

The first approach uses a constraint on the displacement on the two sub-regions $\Gamma_a$ and $\Gamma_b$. Here, we seek $\underline{u}\in \bar{\Omega}_e$ and $u \in \overline{\Omega_\delta \cup \Gamma_a \cup \Gamma_b}$ such that
\begin{equation}
\label{eq:CM-displacement}
\begin{aligned}
- E \uelast''(x) = f_b(x), 
&\quad \forall x \in \Omega_e, \\
- \int_{x-\delta}^{x+\delta} \kappa \frac{\uperid(y) - \uperid(x)}{|y-x|} dy = f_b(x), 
&\quad \forall x \in \Omega_\delta, \\
\uelast(x) = 0, 
&\quad \text{at}\ x = 0, \\
E \uelast'(x) = g, 
&\quad \text{at}\ x = \ell, \\
\uperid(x) - \uelast(x) = 0, 
&\quad \forall x \in \overline{\Gamma_a \cup \Gamma_b}.
\end{aligned}
\end{equation}
A similar approach was proposed in~\cite{DElia-Bochev-2021,zaccariotto2018coupling}. The degree of precision of MDCM is three with respect to the parameter $\delta$, in the sense that any polynomial solution to~\eqref{eq:1dlinearelasticity} of degree up to three will be reproduced exactly by the coupling method~\cite{diehl2022coupling}.

\subsection{Matching-stress coupling method (MSCM)}

In this approach, a constraint on the stresses is provided in the sub-regions $\Gamma_a$ and $\Gamma_b$. Here, we seek $\underline{u}\in \bar{\Omega}_e$ and $u \in \overline{\Omega_\delta \cup \Gamma_a \cup \Gamma_b}$ such that
\begin{equation}
\label{eq:CM-stress}
\begin{aligned}
- E \uelast''(x) = f_b(x), 
&\quad \forall x \in \Omega_e, \\
- \int_{x-\delta}^{x+\delta} \kappa \frac{\uperid(y) - \uperid(x)}{|y-x|} dy = f_b(x), 
&\quad \forall x \in \Omega_\delta, \\
\uelast(x) = 0, 
&\quad \text{at}\ x = 0, \\
E \uelast'(x) = g, 
&\quad \text{at}\ x = \ell, \\
\uperid(x) - \uelast(x) = 0, 
&\quad \text{at}\ x=a, \\
\uperid(x) - \uelast(x) = 0, 
&\quad \text{at}\ x=b, \\
\sigma^+(u)(x) - E\uelast'(x) = 0, 
&\quad \forall x \in \Gamma_a, \\
\sigma^-(u)(x) - E\uelast'(x) = 0, 
&\quad \forall x \in \Gamma_b,
\end{aligned}
\end{equation}
where $\sigma^\pm(u)$ is given by 
\begin{equation}
\begin{aligned}
\label{eq:sigma2}
\sigma^\pm(u)(x)
&= \frac{\delta}{2} 
\int_{x-\delta}^{x} \int_{x}^{z\pm\delta} \kappa \frac{u(y) - u(z)}{|y-z|} dydz \\
&- \frac{\kappa \delta^4}{48} u'''(x).
\end{aligned}
\end{equation}
This approach was inspired by~\cite{silling2020Couplingstresses}. The degree of precision of MSCM is also three as shown in~\cite{diehl2022coupling}. 

\subsection{Coupling method with variable horizon (VHCM)}

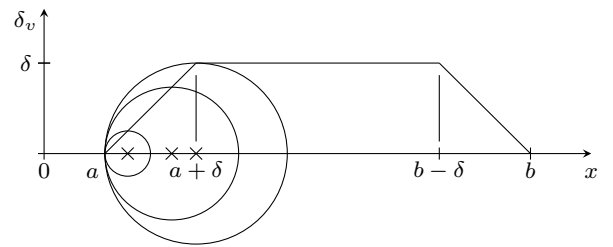
\begin{figure}[htb]
\centering
\begin{tikzpicture}[scale=0.8]
\draw[arrows=->, >=stealth] (-0.1,0) -- (9.0,0);
\draw[arrows=->, >=stealth] (0,-0.1) -- (0,2.4);
\draw (1.0,0.0) -- (2.5,1.5);
\draw (2.5,1.5) -- (6.5,1.5);
\draw (6.5,1.5) -- (8.0,0.0);
\draw (2.5,0.2) -- (2.5,1.3);
\draw (6.5,0.2) -- (6.5,1.3);
\draw (6.5,-0.1) -- (6.5,0.1);
\draw (8.0,-0.1) -- (8.0,0.1);
\draw (-0.1,1.5) -- (0.1,1.5);
\draw (-0.1,1.5) -- (0.1,1.5);
\node at (-0.3,2.2) {$\delta_v$};
\node at (-0.3,1.5) {$\delta$};
\node[above] at (0.0,-0.55) {$0$};
\node[above] at (0.8,-0.55) {$a$};
\node[above] at (2.5,-0.56) {$a+\delta$};
\node[above] at (6.5,-0.56) {$b-\delta$};
\node[above] at (8.0,-0.55) {$b$};
\node[above] at (9.0,-0.55) {$x$};
\draw (2.5,0) circle (1.5);
\draw (2.4,-0.1) -- (2.6,0.1);
\draw (2.4,0.1) -- (2.6,-0.1);
\draw (2.1,0) circle (1.1);
\draw (2.0,-0.1) -- (2.2,0.1);
\draw (2.0,0.1) -- (2.2,-0.1);
\draw (1.375,0) circle (0.375);
\draw (1.275,-0.1) -- (1.475,0.1);
\draw (1.275,0.1) -- (1.475,-0.1);
\end{tikzpicture}
\caption{Example of a variable horizon function $\delta_v(x)$. The circles centered at points $x\in (a,a+\delta)$ are representations of the associated domains $H_\delta(x)$ in terms of $\delta_v(x)$ (adapted from~\cite{diehl2022coupling}).}
\label{Fig:variablehorizon}
\end{figure}

The main motivation when designing the third coupling method is to avoid having to employ overlapping regions between the two models. In order to do so, one can decrease the horizon $\delta$ to zero as one gets closer to the interfaces $x=a$ and $x=b$ while keeping it constant sufficiently far away from them. Let $\delta \in \mathbb{R}^+$ be given. It was proposed in~\cite{Prudhomme-Diehl-2020} that the variable horizon $\delta_v(x)$ be piecewise linear on $\Omega_\delta$, that is
\begin{equation}
\label{eq:deltafn}
\delta_v(x) = \left\{ 
\begin{array}{ll} 
x-a, & \quad a < x \leq a+\delta, \\ 
\delta, & \quad a+\delta < x \leq b - \delta, \\ 
b - x, & \quad b-\delta < x < b,
\end{array}
\right.
\end{equation}
as shown in Figure~\ref{Fig:variablehorizon}. Other choices could be adopted, such as taking $\delta_v$ to be cubic in $[a,a+\delta]$ and $[b-\delta,b]$, but will not be considered in this work. Now, a variable horizon implies that the material parameter $\kappa$ in~\eqref{eq:kappa} also depends on position $x$, even if $E$ is chosen constant, and reads $\kappa(x)=2E/\delta_v(x)^2$. In the general case with varying $E$ and $A$, the parameter would be given by $\kappa(x)=2E(x)A(x)/\delta_v(x)^2$.

The coupling method consists then in searching for $\underline{u}\in \bar{\Omega}_e$ and $u \in \bar{\Omega}_\delta$ such that
\begin{equation}
\label{eq:CM-variablehorizon}
\begin{aligned}
- E \uelast''(x) &= f_b(x), 
\quad \forall x \in \Omega_e, \\
- \int_{x-\delta_v(x)}^{x+\delta_v(x)} \bar\kappa(x) \frac{\uperid(y) - \uperid(x)}{|y-x|} dy &= f_b(x), 
\quad \forall x \in \intdomd, \\
\uelast(x) &= 0, 
\quad \text{at}\ x = 0, \\
E \uelast'(x) &= g, 
\quad \text{at}\ x = \ell, \\
\uperid(x) - \uelast(x) &= 0, 
\quad \text{at}\ x = a, \\
\uperid(x) - \uelast(x) &= 0, 
\quad \text{at}\ x = b, \\
\sigma^+(u)(x) - E\uelast'(x) &= 0, 
\quad \text{at}\ x=a, \\
\sigma^-(u)(x) - E\uelast'(x) &= 0, 
\quad \text{at}\ x=b.
\end{aligned}
\end{equation}
One obvious advantage of the approach over the other two coupling methods is that the overlapping regions $\Gamma_a$ and $\Gamma_b$ are no longer needed, which simplifies to some extent its implementation. \textcolor{blue}{Adaptivity and scaling in one-dimensional PD was introduced in 2009~\cite{https://doi.org/10.1002/nme.2439}}.
A similar coupling approach was proposed in in 2015~\cite{silling2015variable, NIKPAYAM2019308}.

\section{Discretization of the coupling methods with overlapping regions}
\label{sec:coupling:discrete}

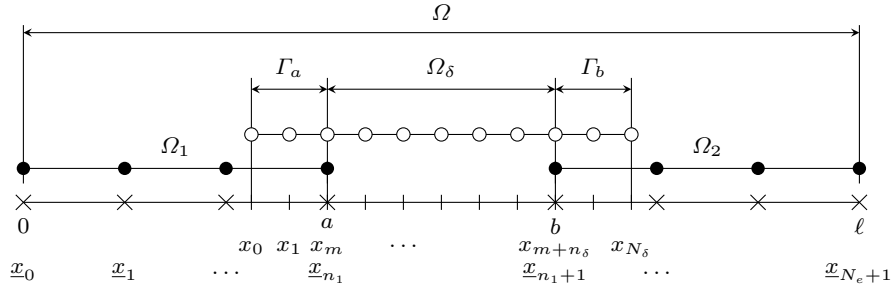
\begin{figure*}[tbp]
\centering
\begin{tikzpicture}
\draw (0,0) -- (11,0);
\foreach \i in {0,...,3}
{\draw (\i*4/3-0.1,-0.1) -- (\i*4/3+0.1,0.1);
\draw (\i*4/3+0.1,-0.1) -- (\i*4/3-0.1,0.1);}
\foreach \i in {0,...,3}
{\draw (7+\i*4/3-0.1,-0.1) -- (7+\i*4/3+0.1,0.1);
\draw (7+\i*4/3+0.1,-0.1) -- (7+\i*4/3-0.1,0.1);}
\foreach \i in {0,...,10}
{\draw (3.0+0.5*\i,-0.1) -- (3.0+0.5*\i,0.1);}
\node[below] at (0.0,-0.7)  {$\underline{x}_0$};
\node[below] at (1.0*4/3,-0.7)  {$\underline{x}_1$};
\node[below] at (2.0*4/3,-0.8)  {$\ldots$};
\node[below] at (4.0,-0.7)  {$\underline{x}_{n_1}$};
\node[below] at (7.0,-0.7)  {$\underline{x}_{n_1+1}$};
\node[below] at (11.0,-0.7)  {$\underline{x}_{N_e+1}$};
\node[below] at (7.0+4/3,-0.8)  {$\ldots$};
\node[below] at (3.0,-0.4)  {$x_0$};
\node[below] at (3.5,-0.4)  {$x_1$};
\node[below] at (4.0,-0.4)  {$x_m$}; 
\node[below] at (7.0,-0.4)  {$x_{m+n_\delta}$}; 
\node[below] at (8.0,-0.4)  {$x_{N_\delta}$};
\node[below] at (5.0,-0.45)  {$\ldots$};
\node[below] at (0.0,-0.1)  {$0$};
\node[below] at (11.0,-0.1)  {$\ell$};
\node[below] at (4.0,-0.1)  {$a$};
\node[below] at (7.0,-0.1)  {$b$};
\draw[arrows=<->, >=stealth] (3.0,1.5) -- (4.0,1.5);
\draw[arrows=<->, >=stealth] (4.0,1.5) -- (7.0,1.5);
\draw[arrows=<->, >=stealth] (7.0,1.5) -- (8.0,1.5);
\draw[arrows=<->, >=stealth] (0.0,2.25) -- (11.0,2.25);
\draw (00.0,0.25) -- (00.0,2.35);
\draw (11.0,0.25) -- (11.0,2.35);
\node[above] at (5.5,2.30) {$\Omega$};
\draw (0.0,0.45) -- (4.0,0.45);
\node[above] at (2.0,0.5) {$\Omega_1$};
\draw (7.0,0.45) -- (11.0,0.45);
\node[above] at (9.0,0.5) {$\Omega_2$};
\draw (3.0,0.9) -- (8.0,0.9);
\node[above] at (3.5,1.55) {$\Gamma_a$};
\node[above] at (5.50,1.55) {$\intdomd$};
\node[above] at (7.5,1.55) {$\Gamma_b$};
\foreach \x in {4,7}
{\draw (\x,0.25) -- (\x,0.55);}
\foreach \x in {3.0,4.0,7.0,8.0}
{\draw (\x,0.0) -- (\x,1.65);}
\foreach \i in {0,...,3}
{\node[circle,color=black,fill=black,inner sep=0pt,minimum size=5pt,label=below:{}] at (\i*4/3,0.45) {};}
\foreach \i in {0,...,10}
{\node[circle,draw=black,fill=white,inner sep=0pt,minimum size=5pt,label=below:{}] at (3.0+0.5*\i,0.90) {};}
\foreach \i in {0,...,3}
{\node[circle,color=black,fill=black,inner sep=0pt,minimum size=5pt,label=below:{}] at (7+\i*4/3,0.45) {};}
\end{tikzpicture}
\caption{Definition of the grid points and degrees of freedom (represented by \(\bullet\) for the degrees of freedom associated with the classical linear elasticity model and by \(\circ\) for the degrees of freedom associated with the peridynamic model) for the coupling methods with overlaps. In this example, $m=2$, $n_\delta = 6$, $N_\delta = 10$, and $n_1=n_2=3$.}
\label{Fig:discretization}
\end{figure*}

We now proceed with the discretization of the coupling methods with overlapping regions described in the previous section. The presentation here will differ from that of~\cite{diehl2022coupling} since the grid sizes for the local and nonlocal models will be chosen different, and the grid points will be possibly non matching. More specifically, we shall consider a different convention for the numbering of the grid points and the nodal degrees of freedom.

\subsection{Construction of the grids}

The first step is to construct the grid in $\Omega_\delta$ that will serve to discretize the peridynamic equation. Let $\delta$ be given and let $m$ be a positive integer, which is often chosen as $m=3$ in the literature. We then introduce the grid size $h_\delta={\delta}/{m}$. We then construct $\Omega_\delta = (a,b)$ in such a way that the length $b-a$ of the domain be a multiple of $h_\delta$, in other words, the number of grid cells in $\Omega_\delta$ is exactly given by $n_\delta=(b-a)/h_\delta$. When considering coupling methods with overlapping regions, one needs to discretize the regions $\Gamma_a$ and $\Gamma_b$ as well, see Figure~\ref{Fig:discretization}. Since the overlapping regions are supposed to be of size $\delta$, each one is thus partitioned into $m$ grid cells. If follows that the total number of grid cells for the domain $\overline{\Gamma_a \cup \Omega_\delta \cup \Gamma_b}$ is given by $N_\delta = n_\delta + 2m$. In other words, the grid points are given by:
\begin{equation}
x_k = (a-\delta) + k h,\quad k=0,1,\ldots,N_\delta.
\end{equation}

The second step is devoted to the discretization of the region~$\Omega_e$. The domain is first decomposed into the two subdomains $\Omega_1 = (0,a)$ and $\Omega_2 = (b,\ell)$, as shown in Figure~\ref{Fig:discretization}. For simplicity, we assume that both domains have the same grid size $h_e$ and that the interface points $a$ and $b$ are such that $\Omega_1$ and $\Omega_2$ can be exactly partitioned into $n_1=a/h_e$ and $n_2=(\ell-b)/h_e$ grid cells, respectively. The grid points in the domains are thus given by
\begin{align}
& \underline{x}_j = j h_e, \quad j=0,1,\ldots,n_1, \\
& \underline{x}_{n_1+1+j} = \ell - (n_2-j) h_e, \quad j=0,1,\ldots,n_2.
\end{align}
The total number of grid cells and of grid points in $\bar{\Omega}_e$ are then equal to $N_e = n_1+n_2$ and $N_e+2$, respectively.

Moreover, we assume that the grids share a grid point at both interfaces, 
\begin{align}
& a = x_m = \underline{x}_{n_1}, \\
& b = x_{m+n_\delta} = \underline{x}_{n_1+1},
\end{align}
and that the grid size for the nonlocal model is smaller than that for the local model, \emph{i.e.}, $h_\delta < h_e$ so that the grid points in the overlapping regions $\Gamma_a$ and $\Gamma_b$ do not necessarily coincide with points in $\Omega_e$. Such an assumption can actually be relaxed, but for the sake of simplicity in the presentation, we suppose that it holds here. Moreover, it is clear that in 2D and 3D, points on the interface will not necessarily coincide. 

\subsection{Degrees of freedom}

The degrees of freedom for the displacements will follow the same numbering as that chosen for the grid points. More specifically, the nodal values for the nonlocal model will be denoted by:
\begin{equation}
u_k \approx u(x_k), \quad k = 0,1,\ldots, N_\delta,
\end{equation}
and the nodal values for the local model by:
\begin{equation}
\underline{u}_j \approx \underline{u}(x_j),\quad  j=0,1,\ldots,N_e+1.
\end{equation}
The total number of degrees of freedom for both the local and nonlocal models is then equal to:
\begin{equation}
N=N_\delta+N_e+3 = n_\delta + 2m +n_1+n_2 +3.
\end{equation}

\subsection{Interpolation operators}

\begin{figure}[tb]
\centering
\begin{tikzpicture}[scale=1.4]
\draw (0,0) -- (5.5,0);
\draw (0,0.45) -- (4,0.45);
\draw (2.0,0.9) -- (5.5,0.9);
\draw (2.0,0.0) -- (2.0,1.65);
\draw (4.0,0.0) -- (4.0,1.65);
\draw[arrows=<->, >=stealth] (2.0,1.5) -- (4.0,1.5);
\foreach \i in {0,...,3}
{\draw (4/3*\i-0.1,-0.1) -- (4/3*\i+0.1,0.1);
\draw (4/3*\i+0.1,-0.1) -- (4/3*\i-0.1,0.1);}
\foreach \i in {0,...,6}
{\draw (2.0+0.5*\i,-0.1) -- (2.0+0.5*\i,0.1);}
\foreach \i in {0,...,3}
{\draw[arrows=<-, >=stealth] (2.0+0.5*\i,0.45) -- (2.0+0.5*\i,0.9);}
    
\node[below] at (0.0,-0.5)  {$\underline{x}_0$};
\node[below] at (4/3,-0.5)  {$\underline{x}_1$};
\node[below] at (8/3,-0.5)  {$\underline{x}_2$};
\node[below] at (4.0,-0.5)  {$\underline{x}_{n_1}$};
\node[below] at (2.0,-0.2)  {$x_0$};
\node[below] at (2.5,-0.2)  {$x_1$};
\node[below] at (3.0,-0.2)  {$x_2$};
\node[below] at (3.5,-0.2)  {$x_3$};
\node[below] at (4.0,-0.2)  {$x_m$};
\node[above] at (1.0,0.6) {$\Omega_1$};
\node[above] at (3.0,1.60) {$\Gamma_a$};
    
\foreach \i in {0,...,3}
{\node[circle,color=black,fill=black,inner sep=0pt,minimum size=5pt,label=below:{}] at (4/3*\i,0.45) {};}

\foreach \i in {0,...,6}
{\node[circle,draw=black,fill=white,inner sep=0pt,minimum size=5pt,label=below:{}] at (2.0+0.5*\i,0.90) {};}
\end{tikzpicture}
\caption{Example of coupling region $\Gamma_a$ where interpolation is needed to match the displacements or stresses at the peridynamic nodes $x_k$, $k=0,\ldots,m-1$, with those from the classical linear elasticity model. In this example, $m=4$ and $n_1=3$.}
\label{fig:interface:interpolation}
\end{figure}
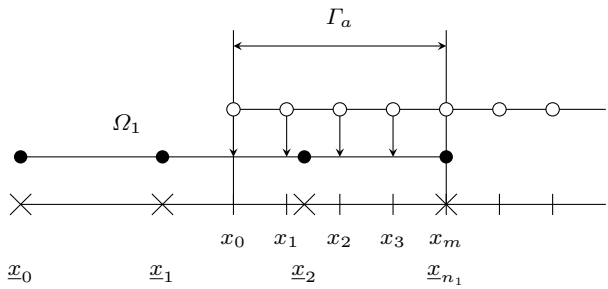

Our goal here is to infer the displacements at the peridynamic nodes $x_k$, $k=0,\ldots,m-1$, in the overlapping region $\Gamma_a$, from the displacement solution $\underline{u}_j$, $j=0,\ldots,n_1$ in $\Omega_1$, see Figure~\ref{fig:interface:interpolation}. Since the grid points are not necessarily matching, a straightforward approach is to use interpolation operators to calculate an interpolating polynomial of the displacements $\underline{u}_j$. In this work, we will use Lagrangian interpolation. We first recall here the general definition of the Lagrange basis for polynomials of degree less or equal to $p$. Given a set of $p+1$ nodal points $\{\underline{x}_0,\underline{x}_1,\ldots,\underline{x}_p\}$, the Lagrange basis polynomial $L_j^p$ of degree~$p$ associated with node $\underline{x}_j$ is given by:
\begin{equation}
L_j^p(x) = \prod_{\substack{i=0 \\ i\neq j}}^p 
\bigg[ \frac{x-\underline{x}_i}{\underline{x}_j-\underline{x}_i} \bigg].
\end{equation}
The Lagrange interpolating polynomial of degree~$p$ pas\-sing through the points $(\underline{x}_j,\underline{u}_j)$, $j=0,\ldots,p$, then reads:
\begin{equation}
I_p(x) = \sum_{j=0}^p \underline{u}_j L_j^p(x).
\end{equation}

Suppose now that we would like to infer the value of $u_{m-1}$ at the node $x_{m-1}$ (\emph{i.e.}, the first peridynamic node to the left of the interface at $x=a$). Since $h_\delta < h_e$, the node $x_{m-1}$ necessarily belongs to the grid cell $[\underline{x}_{n_1-1}, \underline{x}_{n_1}]$. Therefore, the simplest way would be to use the linear interpolating polynomial $I_1$ that passes through the two points $(\underline{x}_{j}, \underline{u}_{j})$, with $j=n_1-1,n_1$. In other words, $u_{m-1}$ would be given by:
\begin{equation}
u_{m-1} 
= I_1 (x_{m-1}) 
= \sum_{j=n_1-1}^{n_1} \underline{u}_{j} L_{j}^1(x_{m-1}),
\end{equation}
where the weighting coefficients associated with the degrees of freedom $\underline{u}_j$ are provided in terms of the two Lagrange basis polynomials $L_{j}^1$, $j=n_1-1$ and $j=n_1$, evaluated at $x=x_{m-1}$:
\begin{equation}
\label{eq:weightsP1}
\begin{aligned}
L_{n_1-1}^1 (x) 
&= \bigg[ \frac{x-\underline{x}_{n_1}}{\underline{x}_{n_1-1}-\underline{x}_{n_1}} \bigg] =
- \frac{(x-\underline{x}_{n_1})}{h_e}, \\
L_{n_1}^1 (x) 
&= \bigg[ \frac{x-\underline{x}_{n_1-1}}{\underline{x}_{n_1}-\underline{x}_{n_1-1}} \bigg] =
\frac{(x-\underline{x}_{n_1-1})}{h_e}.
\end{aligned} 
\end{equation}
Alternatively, one could use the quadratic interpolating polynomial $I_2$ that passes through the three points $(\underline{x}_{j}, \underline{u}_{j})$, with $j=n_1-2, n_1-1, n_1$, in which case one has:
\begin{align}
\label{eq:coupling:md:quadratic}
u_{m-1} 
= I_2 (x_{m-1}) 
= \sum_{j=n_1-2}^{n_1} \underline{u}_{j} L_{j}^2(x_{m-1}),
\end{align}
where the weighting coefficients are provided in terms of the three Lagrange basis polynomials $L_{j}^2$, $j=n_1-2, n_1-1, n_1$, evaluated at $x=x_{m-1}$:
\begin{equation}
\label{eq:weightsP2}
\begin{aligned}
L_{n_1-2}^2 (x) 
&= 
\bigg[ \frac{x-\underline{x}_{n_1-1}}{\underline{x}_{n_1-2}-\underline{x}_{n_1-1}} \bigg] 
\bigg[ \frac{x-\underline{x}_{n_1}}{\underline{x}_{n_1-2}-\underline{x}_{n_1}} \bigg] \\
&= \frac{(x-\underline{x}_{n_1-1})(x-\underline{x}_{n_1})}{2h_e^2},
\\
L_{n_1-1}^2 (x) 
&= 
\bigg[ \frac{x-\underline{x}_{n_1-2}}{\underline{x}_{n_1-1}-\underline{x}_{n_1-2}} \bigg] 
\bigg[ \frac{x-\underline{x}_{n_1}}{\underline{x}_{n_1-1}-\underline{x}_{n_1}} \bigg] \\
&= - \frac{(x-\underline{x}_{n_1-2})(x-\underline{x}_{n_1})}{h_e^2},
\\
L_{n_1}^2 (x) 
&= 
\bigg[ \frac{x-\underline{x}_{n_1-2}}{\underline{x}_{n_1}-\underline{x}_{n_1-2}} \bigg] 
\bigg[ \frac{x-\underline{x}_{n_1-1}}{\underline{x}_{n_1}-\underline{x}_{n_1-1}} \bigg] \\
&= \frac{(x -\underline{x}_{n_1-2})(x -\underline{x}_{n_1-1})}{2h_e^2}.
\end{aligned} 
\end{equation}
Finally, one could also consider the Lagrange interpolating polynomial of degree three, $I_3$, passing through the three points $(\underline{x}_{j}, \underline{u}_{j})$, with $j=n_1-3, \ldots, n_1$, so that
\begin{equation}
\label{eq:mdcm:cubic:interpolation}
u_{m-1} 
= I_3 (x_{m-1}) 
= \sum_{j=n_1-3}^{n_1} \underline{u}_{j} L_{j}^3(x_{m-1}),
\end{equation}
where the weighting coefficients are in this case:
\begin{equation}
\label{eq:weightsP3}
\begin{aligned}
L_{n_1-3}^3 (x) 
&= 
\prod_{\substack{i=n_1-2,n_1-1,n_1}}
\bigg[ \frac{x-\underline{x}_{i}}{\underline{x}_{n_1-3}-\underline{x}_{i}} \bigg]\\
&= - \frac{(x-\underline{x}_{n_1-2})
(x-\underline{x}_{n_1-1})(x-\underline{x}_{n_1})}{6h_e^3},
\\
L_{n_1-2}^3 (x) 
&= 
\prod_{\substack{i=n_1-3,n_1-1,n_1}}
\bigg[ \frac{x-\underline{x}_{i}}{\underline{x}_{n_1-2}-\underline{x}_{i}} \bigg]\\
&= \frac{(x-\underline{x}_{n_1-3})(x-\underline{x}_{n_1-1})(x-\underline{x}_{n_1})}{2h_e^3},
\\
L_{n_1-1}^3 (x) 
&= 
\prod_{\substack{i=n_1-3,n_1-2,n_1}}
\bigg[ \frac{x-\underline{x}_{i}}{\underline{x}_{n_1-1}-\underline{x}_{i}} \bigg]\\
&= - \frac{(x-\underline{x}_{n_1-3})(x-\underline{x}_{n_1-2})(x-\underline{x}_{n_1})}{2h_e^3},
\\
L_{n_1}^3 (x) 
&= 
\prod_{\substack{i=n_1-3,n_1-2,n_1-1}}
\bigg[ \frac{x-\underline{x}_{i}}{\underline{x}_{n_1}-\underline{x}_{i}} \bigg] \\
&= \frac{(x-\underline{x}_{n_1-3})(x-\underline{x}_{n_1-2})(x-\underline{x}_{n_1-1})}{6h_e^3} .
\end{aligned} 
\end{equation}

In the configuration of Figure~\ref{fig:interface:interpolation}, we observe that all peridynamic nodes in $\bar{\Gamma}_a$ actually belong to the interval $[\underline{x}_{n_1-2}, \underline{x}_{n_1}]$. In other words, one can use the same quadratic interpolant polynomial $I_2$ to evaluate the displacements $u_k$, $k=0,\ldots,m$ (including the point $x_m=a$), that is:
\begin{equation}
u_k = I_2(x_k), \quad k=0,1,\ldots,m.
\end{equation}
The same remark also holds in the case of $I_3$, \emph{i.e.},
\begin{equation}
u_k = I_3(x_k), \quad k=0,1,\ldots,m.
\end{equation}
On the other hands, if one wants to use the linear interpolating polynomial, we see that the situation is a little more complex. Only the points $x_k$, $k=m-1$, $m$, belongs to $[\underline{x}_{n_1-1}, \underline{x}_{n_1}]$. In the configuration of the figure, the points $x_0$ and $x_1$ belongs to the interval $[\underline{x}_{n_1-2}, \underline{x}_{n_1-1}]$. In that case, one needs to define the interpolant of degree one that passes through the two points $(\underline{x}_{n_1-2}, \underline{u}_{n_1-2})$ and $(\underline{x}_{n_1-1}, \underline{u}_{n_1-1})$. For the sake of simplicity in the notation, we will simply denote the new interpolant by $I_1$, as before, and write:
\begin{equation}
u_k = I_1(x_k), \quad k=0,1,\ldots,m,
\end{equation}
understanding that the definition of $I_1$ may change depending on which interval the peridynamic node $x_k$ belongs to.

We shall use in the numerical simulations the Python package \texttt{SymPy}~\cite{10.7717/peerj-cs.103} to evaluate the Lagrange basis polynomials $L_j^p$ at the points $x_k$. We will show in particular the influence of the degree of the interpolating polynomial on the results of the coupling methods.

\subsection{Discretization of MDCM}

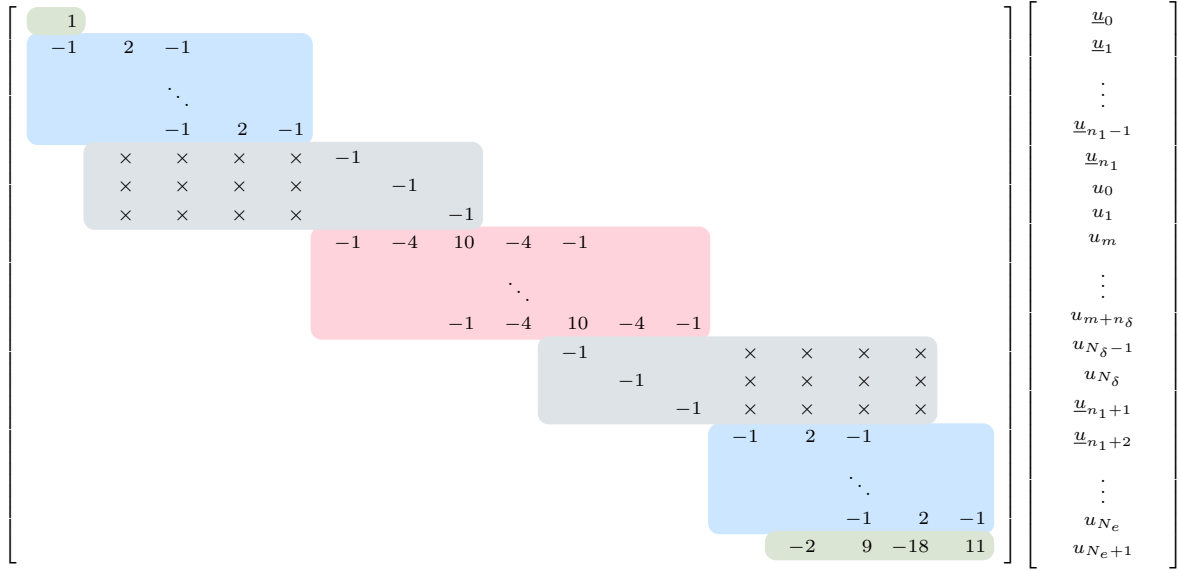
\begin{figure*}[tb]
\centering
\scriptsize
\begin{tikzpicture}
\matrix (m)[
    matrix of math nodes, 
    nodes in empty cells,
    nodes={text width={width(9999)}, 
    align=right},
    right delimiter=\rbrack,left delimiter=\lbrack
  ] {
1 \\
-1 & 2 & -1 & & \\
& & \ddots & & & & \\
& & -1 & 2 & -1 \\
& \times & \times & \times & \times & -1 & & & \\
& \times & \times & \times & \times & & -1 & & & \\
& \times & \times & \times & \times & & & -1 & & \\
& & & & & -1 & -4 & 10 & -4 & -1 \\
& & & & & & & & \ddots & \\
& & & & & & & -1 & -4 & 10 & -4 & -1 \\
& & & & & & & & & -1 & & & \times & \times & \times & \times \\
& & & & & & & & & & -1 & & \times & \times & \times & \times \\
& & & & & & & & & & & -1 & \times & \times & \times & \times \\
& & & & & & & & & & & & -1 & 2 & -1 \\
& & & & & & & & & & & & & & \ddots \\
& & & & & & & & & & & & & & -1 & 2 & -1 \\
& & & & & & & & & & & & & -2 & 9 & -18 & 11 \\
} ;
\begin{pgfonlayer}{myback}
\fhighlight[azure!20]{m-2-1}{m-4-5}
\fhighlightL[azure!20]{m-13-13}{m-16-17}
\fhighlightL[awesome!20]{m-7-6}{m-10-12}
\fhighlightL[cadetgrey!30]{m-4-2}{m-7-8}
\fhighlightL[cadetgrey!30]{m-10-10}{m-13-16}
\fhighlight[asparagus!30]{m-1-1}{m-1-1}
\fhighlightL[asparagus!30]{m-16-14}{m-17-17}
\end{pgfonlayer}
\matrix (r)[
    matrix of math nodes,
    nodes in empty cells,
    nodes={text width={width(999999999)}, 
    align=center},
    right delimiter=\rbrack,left delimiter=\lbrack
  ] at (7.8,0) {
\underline{u}_0 \\
\underline{u}_1 \\
\vdots \\
\underline{u}_{n_1-1} \\
\underline{u}_{n_1} \\
u_0 \\
u_1 \\
u_m \\
\vdots\\
u_{m+n_\delta} \\
u_{N_\delta-1} \\
u_{N_\delta} \\
\underline{u}_{n_1+1} \\
\underline{u}_{n_1+2} \\
\vdots\\
u_{N_e} \\
u_{N_e+1} \\
} ;
\end{tikzpicture}
\caption{Sketch of the assembled stiffness matrix for the coupling method with matching stresses (MDCM). The first and the last rows, shown in {green}, correspond to the Dirichlet boundary condition at $\underline{x}_0$ and the Neumann boundary condition at $\underline{x}_{N_e+1}$, respectively. The two blocks in {blue} correspond to $\Omega_1$ and $\Omega_2$ while the block in {red} corresponds to $\Omega_\delta$. Note that, for the sake of simplicity, the zero entries are not shown and the parts of the coefficients involving $E$ or $\kappa\delta^2/2$ and the denominators $h_e^2$, $6h_e$, or $8h_\delta^2$ are omitted. The two blocks in {grey} correspond to the overlapping regions $\Gamma_a$ and $\Gamma_b$. Here, we consider $m=2$ and the interpolant $I_3$ in Equation~\eqref{eq:mdcm:cubic:interpolation}. The symbols $\times$ in the top grey box correspond to the weighting coefficients~\eqref{eq:weightsP3} evaluated at the points $x_{m-2}$, $x_{m-1}$, and $x_m$ associated with the degrees of freedom $u_{m-2}$, $u_{m_1}$, and $u_m$, respectively. The same comment holds for the bottom grey box.}
\label{fig:MDCMmatrix}
\end{figure*}

The second derivative appearing in the classical linear elasticity model in~\eqref{eq:CM-displacement} is approximated by the second-order central difference stencil. For the discretization of the integral in~\eqref{eq:CM-displacement}, a classical second-order trapezoidal integration rule is used. Here, more advanced quadrature rules or quadrature rules specific to non-local models could be used. For more details, we refer to~\cite{Bilodeau2024}. For the Neumann condition, we use a one-sided third-order finite difference formula so as to ensure that the approximation is error-free when the solution is a polynomial function of degree at most three. The discretization of Problem~\eqref{eq:CM-displacement} leads to the following system of equations:
\begin{enumerate}[itemsep=0pt,topsep=4pt,parsep=0pt,leftmargin=13pt]
\item 
Dirichlet BC at \(x=\underline{x}_0=0\):
\begin{equation}
\underline{u}_0 = 0.
\end{equation}
\item 
In \(\Omega_1\): 
For \(j=1,\ldots,n_1-1\),
\begin{equation}
\label{eq:discreteCLE1}
- E \frac{\underline{u}_{j-1}-2\underline{u}_j+\underline{u}_{j+1}}{h_e^2} = f_b(\underline{x}_j).
\end{equation}
\item
In \(\bar{\Gamma}_a\): 
For \(k=0,\ldots,m\),
\begin{equation}
\label{eq:couplingeqMDCMa}
I_p(x_{k}) - u_{k} = 0.
\end{equation}
\item 
In \(\bar{\Omega}_\delta\): 
For \( k=m,\ldots,m+n_\delta\),
\begin{equation}
\label{eq:mdcm:pd:constant}
- \frac{\kappa\delta^2}{2} \frac{u_{k-2}+4u_{k-1}-10u_k+4u_{k+1}+u_{k+2}}{8h_\delta^2} = f_b(x_k).
\end{equation}
\item 
In \(\bar{\Gamma}_b\): 
For \(k=m+n_\delta,\ldots,N_\delta\),
\begin{equation}
\label{eq:couplingeqMDCMb}
I_p(x_{k}) - u_{k} = 0.
\end{equation}
\item 
In \(\Omega_2\): 
For \(j=n_1+2,\ldots,N_e\),
\begin{equation}
\label{eq:discreteCLE2}
- E \frac{\underline{u}_{j-1}-2\underline{u}_j+\underline{u}_{j+1}}{h_e^2} = f_b(\underline{x}_j).
\end{equation}
\item 
Neumann BC at \(x=\underline{x}_{N_e+1}=\ell\):
\begin{equation}
\label{eq:discreteNeumannBC}
E \frac{-2\underline{u}_{N_e-2}+9\underline{u}_{N_e-1}-18\underline{u}_{N_e}+11\underline{u}_{N_e+1}}{6h_e} = g.
\end{equation}
\end{enumerate}
We observe that above set of equations exactly provides a total of $N=n_\delta+2m+n_1+n_2+3$ equations in order to solve for the $N$ degrees of freedom. Figure~\ref{fig:MDCMmatrix} shows the resulting matrix associated with the coupled system. 

\begin{remark}
In the case of a the modulus of elasticity~$E=E(x)$ is chosen non-constant, as in~\eqref{eq:1dlinearelasticity},  Equations~\eqref{eq:discreteCLE1} and~\eqref{eq:discreteCLE2} should be replaced by:
\[
- \frac{
  E_{j-1/2} (\underline{u}_{j-1} - \underline{u}_j) 
- E_{j+1/2} (\underline{u}_j - \underline{u}_{j+1})}
{h_e^2} = f_b(\underline{x}_j).
\]
where $E_{j \pm 1/2} = E(\underline{x}_j \pm h_e/2)$. Similarly, the discrete Neumann condition~\eqref{eq:discreteNeumannBC} should then read:
\[
E_{N_e+1} \frac{-2\underline{u}_{N_e-2}+9\underline{u}_{N_e-1}-18\underline{u}_{N_e}+11\underline{u}_{N_e+1}}{6h_e} = g.
\]
Accordingly, the peridynamic stiffness constant $\kappa$ would now vary with $x$ so that~\eqref{eq:mdcm:pd:constant} should be changed to:
\[ 
\begin{aligned}
- \frac{\delta^2}{16h_\delta^2} 
& \Big[ \kappa_{k-2} u_{k-2} +  4 \kappa_{k-1} u_{k-1}
\\
& - \big( \kappa_{k-2} + 4 \kappa_{k-1} 
  + 4 \kappa_{k+1} + \kappa_{k+2} \big) u_k  
\\ 
& + 4 \kappa_{k+1} u_{k+1} + \kappa_{k+2} u_{k+2} 
\Big] = f_b(x_k),
\end{aligned}
\]
where $\kappa_{k\pm 1} = \kappa(x_{k\pm 1})$ and $\kappa_{k\pm 2} = \kappa(x_{k\pm 2})$.
\end{remark}

\subsection{Discretization of MSCM}

We now proceed with the discretization of the coupled system~\eqref{eq:CM-stress}. We keep here the same grid structure and numbering of the nodal points and degrees of freedom and construct the same interpolation operators as for MDCM. The only equations that need to be modified are the coupling relations~\eqref{eq:couplingeqMDCMa} and~\eqref{eq:couplingeqMDCMb} in $\bar{\Gamma}_a$ and $\bar{\Gamma}_b$, respectively. 

At the points $x=a=x_m$ and $x=b=x_{m+n_\delta}$, we require that $\underline{u}_{n_1}=u_m$ and $\underline{u}_{n_1+1}=u_{m+n_\delta}$, or equivalently:
\begin{equation}
\begin{aligned}
&I_p(x_m) - u_m = 0, \\
&I_p(x_{m+n_\delta}) - u_{m+n_\delta} = 0.
\end{aligned}
\end{equation}

\begin{figure*}[tb]
\centering
\scriptsize
\begin{tikzpicture}
\matrix (m)[
    matrix of math nodes, 
    nodes in empty cells,
    nodes={text width={width(9999)}, 
    align=right},
    right delimiter=\rbrack,left delimiter=\lbrack
  ] {
1 \\
-1 & 2 & -1 & & \\
& & \ddots & & & & \\
& & -1 & 2 & -1 \\
& \otimes & \otimes & \otimes & \otimes & 11 & -18 & 9 & -2\\
& \otimes & \otimes & \otimes & \otimes & & 11 & -18 & 9 & -2 \\
& \times & \times & \times & \times & & & -1 & & \\
& & & & & -1 & -4 & 10 & -4 & -1 \\
& & & & & & & & \ddots & \\
& & & & & & & -1 & -4 & 10 & -4 & -1 \\
& & & & & & & & & -1 & & & \times & \times & \times & \times \\
& & & & & & & 2 & -9 & 18 & -11 & & \otimes & \otimes & \otimes & \otimes \\
& & & & & & & & 2 & -9 & 18 & -11 & \otimes & \otimes & \otimes & \otimes \\
& & & & & & & & & & & & -1 & 2 & -1 \\
& & & & & & & & & & & & & & \ddots \\
& & & & & & & & & & & & & & -1 & 2 & -1 \\
& & & & & & & & & & & & & -2 & 9 & -18 & 11 \\
} ;
\begin{pgfonlayer}{myback}
\fhighlight[azure!20]{m-2-1}{m-4-5}
\fhighlightL[azure!20]{m-13-13}{m-16-17}
\fhighlightL[awesome!20]{m-7-6}{m-10-12}
\fhighlightL[cadetgrey!30]{m-4-2}{m-7-10}
\fhighlightL[cadetgrey!30]{m-10-8}{m-13-16}
\fhighlight[asparagus!30]{m-1-1}{m-1-1}
\fhighlightL[asparagus!30]{m-16-14}{m-17-17}
\end{pgfonlayer}
\matrix (r)[
    matrix of math nodes,
    nodes in empty cells,
    nodes={text width={width(999999999)}, 
    align=center},
    right delimiter=\rbrack,left delimiter=\lbrack
  ] at (7.8,0) {
\underline{u}_0 \\
\underline{u}_1 \\
\vdots \\
\underline{u}_{n_1-1} \\
\underline{u}_{n_1} \\
u_0 \\
u_1 \\
u_m \\
\vdots\\
u_{m+n_\delta} \\
u_{N_\delta-1} \\
u_{N_\delta} \\
\underline{u}_{n_1+1} \\
\underline{u}_{n_1+2} \\
\vdots\\
u_{N_e} \\
u_{N_e+1} \\
} ;
\end{tikzpicture}
\caption{Sketch of the assembled stiffness matrix for the coupling method with matching stresses (MSCM). See the caption of Figure~\ref{fig:MDCMmatrix} for a detailed  description of the matrix. Again, the two blocks in {grey} correspond to the overlapping regions $\Gamma_a$ and $\Gamma_b$ using $m=2$ and the interpolant $I_3$ in Equation~\eqref{eq:mdcm:cubic:interpolation}. On one hand, the symbols $\times$ in the grey boxes correspond to the weighting coefficients~\eqref{eq:weightsP3} evaluated at the points $x_m$ and $x_{m+n_\delta}$ associated with the degrees of freedom $u_m$ and $u_{m+n_\delta}$, respectively. On the other hand, the symbols $\otimes$ denote the weighting coefficients in the expression of the derivative of $I_3$ evaluated at the points $x_{m_2}$, $x_{m-1}$, $x_{N_\delta-1}$, and $x_{N_\delta}$.}
\label{fig:MSCMmatrix}
\end{figure*}
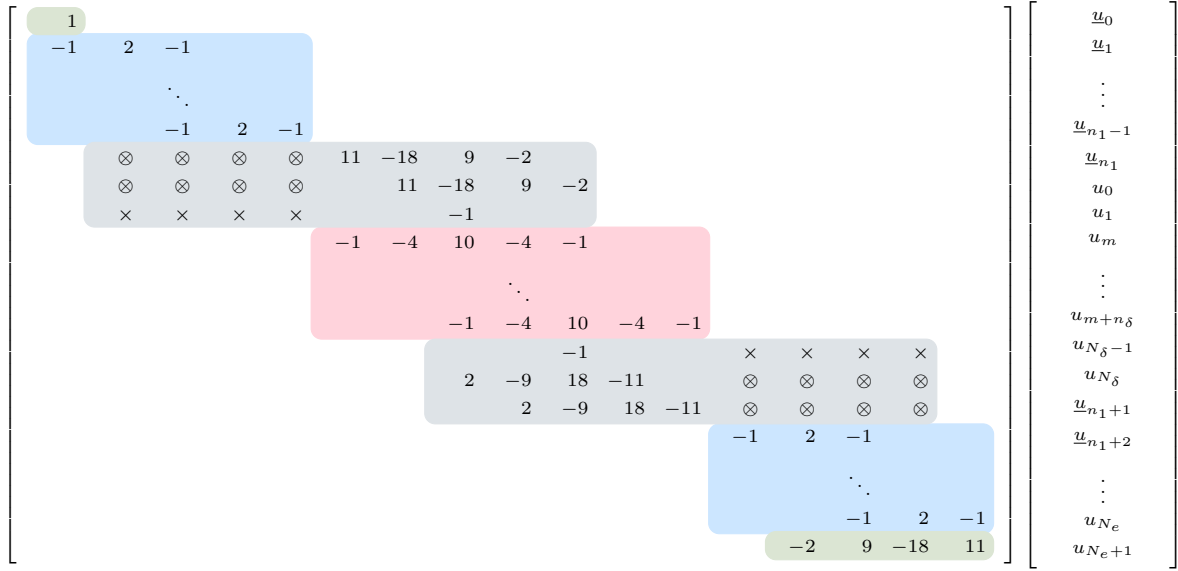

For the constraints on stresses in $\Gamma_a$ and $\Gamma_b$, we shall consider the following coupling equations:
\begin{equation}
\begin{aligned}
&E \frac{dI_p}{dx}(x_k) - \sigma_h^+(x_k) = 0, \quad k=0,\ldots, m-1, \\
&E \frac{dI_p}{dx}(x_k) - \sigma_h^-(x_k) = 0, \quad 
k=m+1+n_\delta, \ldots, N_\delta,
\end{aligned}
\end{equation}
where:
\[
\frac{dI_p}{dx}(x_k) = \sum_{j} \underline{u}_j \frac{L_j^p}{dx}(x_k),
\]
with $p=1$, $2$, or $3$, and the stresses $\sigma_h^{\pm}$ are approximated by one-sided third-order finite differences stencils of the first derivative, see~\cite{diehl2022coupling}:
\begin{equation}
\label{eq:stressapprox}
\begin{aligned}
&\sigma_h^{+}(x_k) = 
\frac{\kappa \delta^2}{2}
\frac{-11u_k + 18u_{k+1} - 9u_{k+2} + 2u_{k+3}}{6h_\delta},
\\
&\sigma_h^{-}(x_k) = 
\frac{\kappa \delta^2}{2}
\frac{-2u_{k-3} + 9u_{k-2} - 18u_{k-1} + 11u_k}{6h_\delta}.
\end{aligned}
\end{equation}
The structure of the resulting stiffness matrix is shown in Figure~\ref{fig:MSCMmatrix}.

\begin{remark}
We note here that the proposed discretization of MSCM is slightly different from the one presented in~\cite{diehl2022coupling}. The peridynamic equation is actually solved at the points $x=a$ and $x=b$ rather than applying the constraint on the stress at those points. The two formulations actually lead to the same results. The structures of the stiffness matrices for MDCM and MSCM shown in Figures~\ref{fig:MDCMmatrix} and~\ref{fig:MSCMmatrix} are then similar.
\end{remark}

\begin{remark}
In the case that the points $x_m$ and $\underline{x}_{n_1}$ (resp.\ $x_{m+n_\delta}$ and $\underline{x}_{n_1}$) do not coincide at $x=a$ (resp.\ $x=b$), that is, the grid for the peridynamic model is constructed in a such that the node $x_m$ (resp.\ $x_{m+n_\delta}$) is slightly moved to the left of $x=a$ (resp.\ to the right of $x=b$), the stiffness matrix would still be assembled as in Figures~\ref{fig:MDCMmatrix} and~\ref{fig:MSCMmatrix} in order to solve for the coupled solutions. 
\end{remark}

\section{Discretization of VHCM}

One advantage of VHCM over the other two coupling methods, MDCM and MSCM, is that there is no overlapping regions when we assume that the end points of the local and nonlocal models exactly coincide. In that configuration, the coupling system would be exactly the same as the one provided in~\cite{diehl2022coupling}, even if the grid sizes $h_\delta$ and $h_e$ take different values. 

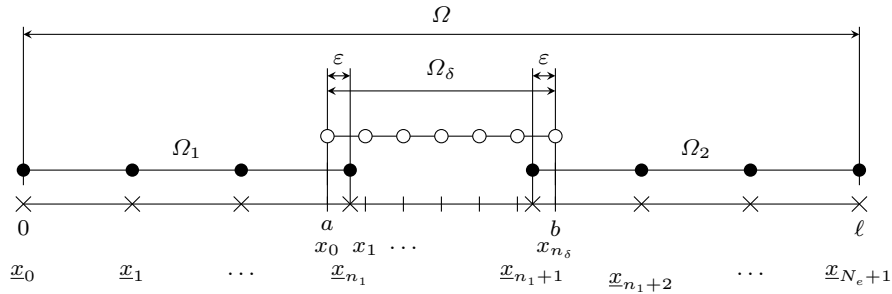
\begin{figure*}[tbp]
\centering
\begin{tikzpicture}
\draw (0,0) -- (11,0);
\foreach \i in {0,...,3}
{\draw (\i*4/3+\i*0.1-0.1,-0.1) -- (\i*4/3+\i*0.1+0.1,0.1);
\draw (\i*4/3+0.1+\i*0.1,-0.1) -- (\i*4/3-0.1+\i*0.1,0.1);}
\foreach \i in {0,...,3}
{\draw (7+\i*4/3-0.1+\i*0.1-0.3,-0.1) -- (7+\i*4/3+0.1+\i*0.1-0.3,0.1);
\draw (7+\i*4/3+0.1+\i*0.1-0.3,-0.1) -- (7+\i*4/3-0.1+\i*0.1-0.3,0.1);}
\foreach \i in {0,...,6}
{\draw (4.0+0.5*\i,-0.1) -- (4.0+0.5*\i,0.1);}
\node[below] at (0.0,-0.7)  {$\underline{x}_0$};
\node[below] at (1.0*4/3+0.1,-0.7)  {$\underline{x}_1$};
\node[below] at (2.0*4/3+0.2,-0.8)  {$\ldots$};
\node[below] at (4.0+0.3,-0.7)  {$\underline{x}_{n_1}$};
\node[below] at (7.0-0.3,-0.7)  {$\underline{x}_{n_1+1}$};
\node[below] at (11.0,-0.7)  {$\underline{x}_{N_e+1}$};
\node[below] at (7.0+4/3-0.2,-0.8)  {$\underline{x}_{n_1+2}$};
\node[below] at (7.0+2*4/3-0.1,-0.8)  {$\ldots$};
\node[below] at (4.0,-0.4)  {$x_0$}; 
\node[below] at (4.5,-0.4)  {$x_1$}; 
\node[below] at (7.0,-0.4)  {$x_{n_\delta}$}; 
\node[below] at (5.0,-0.45)  {$\ldots$};
\node[below] at (0.0,-0.1)  {$0$};
\node[below] at (11.0,-0.1)  {$\ell$};
\node[below] at (4.0,-0.1)  {$a$};
\node[below] at (7.0,-0.1)  {$b$};
\draw[arrows=<->, >=stealth] (4.0,1.7) -- (4.3,1.7);
\draw[arrows=<->, >=stealth] (4.0,1.5) -- (7.0,1.5);
\draw[arrows=<->, >=stealth] (6.7,1.7) -- (7.0,1.7);
\draw[arrows=<->, >=stealth] (0.0,2.25) -- (11.0,2.25);
\draw (00.0,0.25) -- (00.0,2.35);
\draw (11.0,0.25) -- (11.0,2.35);
\node[above] at (5.5,2.30) {$\Omega$};
\draw (0.0,0.45) -- (4.3,0.45);
\node[above] at (2.15,0.5) {$\Omega_1$};
\draw (7.0-0.3,0.45) -- (11.0,0.45);
\node[above] at (8.85,0.5) {$\Omega_2$};
\draw (4.0,0.9) -- (7.0,0.9);
\node[above] at (4.15,1.75) {$\varepsilon$};
\node[above] at (5.50,1.55) {$\intdomd$};
\node[above] at (6.85,1.75) {$\varepsilon$};
\foreach \x in {4,7}
{\draw (\x,0.25) -- (\x,0.55);}
\foreach \x in {4.0,4.3,6.7,7.0}
{\draw (\x,0.0) -- (\x,1.80);}
\foreach \i in {0,...,3}
{\node[circle,color=black,fill=black,inner sep=0pt,minimum size=5pt,label=below:{}] at (\i*4/3+\i*0.1,0.45) {};}
\foreach \i in {0,...,6}
{\node[circle,draw=black,fill=white,inner sep=0pt,minimum size=5pt,label=below:{}] at (4.0+0.5*\i,0.90) {};}
\foreach \i in {0,...,3}
{\node[circle,color=black,fill=black,inner sep=0pt,minimum size=5pt,label=below:{}] at (7+\i*4/3+\i*0.1-0.3,0.45) {};}
\end{tikzpicture}
\caption{Definition of the grid points and degrees of freedom (represented by \(\bullet\) for the degrees of freedom associated with the classical linear elasticity model and by \(\circ\) for the degrees of freedom associated with the peridynamic model) for VHCM. In this example, $n_\delta = 6$, and $n_1=n_2=3$.}
\label{Fig:discretizationVHCM}
\end{figure*}

In this section, we relax this assumption and suppose that the two grids do not necessarily match at their extremities, in other words, we consider that the grids for the classical linear elasticity model slightly overlaps the peridynamic grid by an offset of $\varepsilon$, as shown in Figure~\ref{Fig:discretizationVHCM}. We will suppose here that the offset $\varepsilon$ is smaller than the grid size used in the nonlocal model. 

The grids are now constructed as follows. As before, the peridynamic domain $\Omega_\delta = (a,b)$ is decomposed into $n_\delta=(b-a)/h_\delta$ cells of size $h_\delta={\delta}/{m}$. The grid points are then given by:
\begin{equation}
x_k = a + k h,\quad k=0,1,\ldots,n_\delta.
\end{equation}
Now, the domains $\Omega_1=(0,a+\varepsilon)$ and $\Omega_2=(b-\varepsilon,\ell)$ are partitioned into $n_1=(a+\varepsilon)/h_e$ and $n_2=(\ell-b-\varepsilon)/h_e$ grid cells, respectively, with grid size $h_e$. The grid points in these domains are thus given by:
\begin{align}
& \underline{x}_j = j h_e, \quad j=0,1,\ldots,n_1, \\
& \underline{x}_{n_1+1+j} = \ell - (n_2-j) h_e, \quad j=0,1,\ldots,n_2,
\end{align}
and the total number of grid cells in $\bar{\Omega}_e$ is thus $N_e = n_1+n_2$. 

As before, we use the same numbering for the degrees of freedom associated with the displacements, \emph{i.e.}, the nodal values for the nonlocal model will be denoted by:
\begin{equation}
u_k \approx u(x_k), \quad k = 0,1,\ldots, n_\delta
\end{equation}
and the nodal values for the local model by:
\begin{equation}
\underline{u}_j \approx \underline{u}(x_j),\quad  j=0,1,\ldots,N_e+1.
\end{equation}
The total number of degrees of freedom is in this case equal to $N=n_\delta+N_e+3 = n_\delta+n_1+n_2+3$.

Proceeding in a similar manner as for MDCM and MSCM, the discretization of VHCM leads to the following system of equations (with $m=2$):
\begin{enumerate}[itemsep=0pt,topsep=4pt,parsep=0pt,leftmargin=13pt]
\item 
Dirichlet BC at \(x=\underline{x}_0=0\):
\begin{equation}
\underline{u}_0 = 0.
\end{equation}
\item 
In \(\Omega_1\): 
For \( j=1,\ldots,n_1-1\),
\begin{equation}
\label{eq:VHCM-CLE1}
- E \frac{\underline{u}_{j-1}-2\underline{u}_j+\underline{u}_{j+1}}{h_e^2} = f_b(\underline{x}_j).
\end{equation}
\item
At interface $x=a$:
\begin{equation}
\label{eq:couplingeqVHCMa}
\begin{aligned}
&I_p(x_0) - u_0 = 0,\\
&E \frac{dI_p}{dx}(x_0) - \sigma_h^+(x_0) = 0. 
\end{aligned}
\end{equation}
\item 
In \(\Omega_\delta\): 
\begin{equation}
\label{eq:VHCM-peridynamics}
\begin{aligned}
&- \frac{\kappa\delta^2}{2} 
\frac{u_{0}-2u_{1}+u_{2}}{h_\delta^2} = f_b(x_1),\\
&- \frac{\kappa\delta^2}{2} \frac{u_{k-2}+4u_{k-1}-10u_k+4u_{k+1}+u_{k+2}}{8h_\delta^2} = f_b(x_k),\\
&\qquad\qquad k=2,\ldots,n_\delta-2,\\
&- \frac{\kappa\delta^2}{2} 
\frac{u_{n_\delta-2}-2u_{n_\delta-1}+u_{n_\delta}}{h_\delta^2} = f_b(x_{n_\delta-1}).
\end{aligned}
\end{equation}
\item 
At interface $x=b$: 
\begin{equation}
\label{eq:couplingeqVHCMb}
\begin{aligned}
&I_p(x_{n_\delta}) - u_{n_\delta} = 0,\\
&E \frac{dI_p}{dx}(x_{n_\delta}) - \sigma_h^-(x_{n_\delta}) = 0.
\end{aligned}
\end{equation}
\item 
In \(\Omega_2\): 
For \(j=n_1+2,\ldots,N_e\),
\begin{equation}
- E \frac{\underline{u}_{j-1}-2\underline{u}_j+\underline{u}_{j+1}}{h_e^2} = f_b(\underline{x}_j).
\end{equation}
\item 
Neumann BC at \(x=\underline{x}_{N_e+1}=\ell\):
\begin{equation}
E \frac{-2\underline{u}_{N_e-2}+9\underline{u}_{N_e-1}-18\underline{u}_{N_e}+11\underline{u}_{N_e+1}}{6h_e} = g.
\end{equation}
\end{enumerate}

\begin{figure*}[tb]
\centering
\scriptsize
\begin{tikzpicture}
  \matrix (m)[
    matrix of math nodes,
    nodes in empty cells,
    nodes={text width={width(9999)}, 
    align=right},
    right delimiter=\rbrack,left delimiter=\lbrack
  ] {
1 \\
-1 & 2 & -1 & & \\
& & \ddots \\
& & -1 & 2 & -1 \\
& \times & \times & \times & \times & -1 \\
& \otimes & \otimes & \otimes & \otimes & 11 & -18 & 9 & -2 \\
& & & & & -1 & 2 & -1 & \\
& & & & & -1 & -4 & 10 & -4 & -1 \\
& & & & & & & & \ddots & \\
& & & & & & & -1 & -4 & 10 & -4 & -1 \\
& & & & & & & & & -1 & 2 & -1 \\
& & & & & & & & 2 & -9 & 18 & -11 & \otimes & \otimes & \otimes & \otimes \\
& & & & & & & & & & & -1 & \times & \times & \times & \times \\
& & & & & & & & & & & & -1 & 2 & -1 & & \\
& & & & & & & & & & & & & & \ddots & & \\
& & & & & & & & & & & & & & -1 & 2 & -1 \\
& & & & & & & & & & & & & -2 & 9 & -18 & 11 \\
} ;
\begin{pgfonlayer}{myback}
\fhighlight[azure!20]{m-2-1}{m-4-5}
\fhighlight[azure!20]{m-14-13}{m-16-17}
\fhighlight[awesome!20]{m-7-6}{m-11-12}
\fhighlightL[cadetgrey!30]{m-4-2}{m-6-9}
\fhighlightL[cadetgrey!30]{m-11-9}{m-13-16}
\fhighlight[asparagus!30]{m-1-1}{m-1-1}
\fhighlight[asparagus!30]{m-17-14}{m-17-17}
\end{pgfonlayer}
\matrix (r)[
    matrix of math nodes,
    nodes in empty cells,
    nodes={text width={width(999999999)}, 
    align=center},
    right delimiter=\rbrack,left delimiter=\lbrack
  ] at (7.8,0) {
\underline{u}_0 \\
\underline{u}_1 \\
\vdots \\
\underline{u}_{n_1-1} \\
\underline{u}_{n_1} \\
u_0 \\
u_1 \\
u_2 \\
\vdots\\
u_{n_\delta-2} \\
u_{n_\delta-1} \\
u_{n_\delta} \\
\underline{u}_{n_1+1} \\
\underline{u}_{n_1+2} \\
\vdots\\
u_{N_e} \\
u_{N_e+1} \\
} ;
\end{tikzpicture}
\caption{Sketch of the assembled stiffness matrix for the coupling method with variable horizon (VHCM). See caption of Figure~\ref{fig:MSCMmatrix} for a detailed  description of the matrix. The two blocks in {grey} correspond to the constraints applied at $x=a$ and $x=b$ using the interpolant $I_3$ in Equation~\eqref{eq:mdcm:cubic:interpolation}. The symbols $\times$ in the grey boxes correspond here to the weighting coefficients~\eqref{eq:weightsP3} evaluated at $x_0$ and $x_{n_\delta}$ associated with the degrees of freedom $u_0$ and $u_{n_\delta}$, respectively. The symbols $\otimes$ denote the weighting coefficients in the expression of the derivative of $I_3$ evaluated at the points $x_{0}$ and $x_{n_\delta}$.}
\label{fig:VHCMmatrix}
\end{figure*}
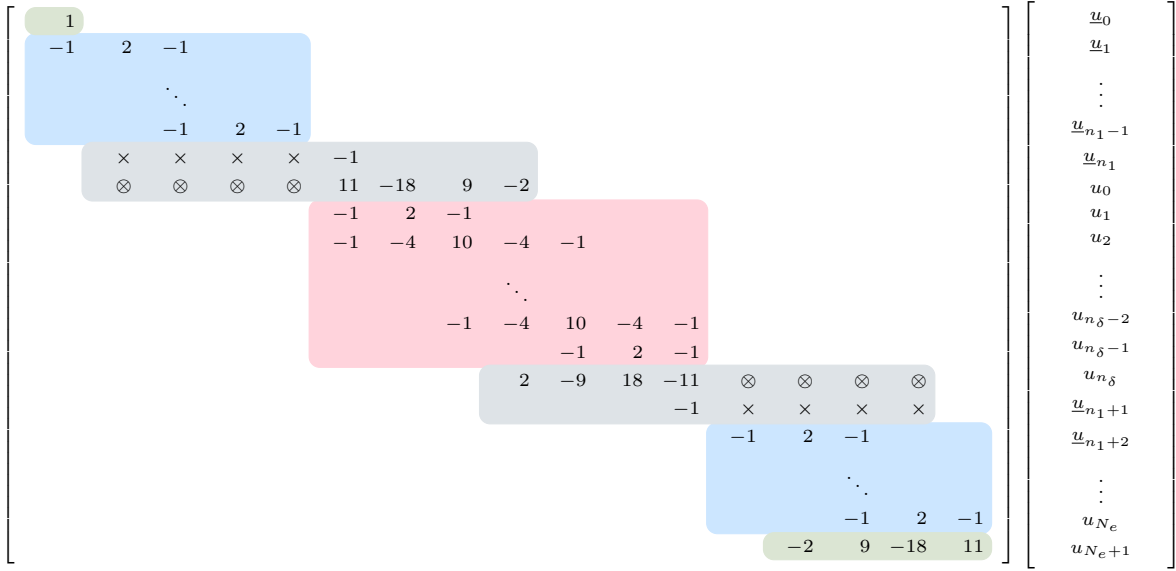

In the above, we recall that if $E$ is constant, then the quantity $\bar{\kappa}(x)\delta_v(x)^2 = \kappa\delta^2$ remains also constant. Moreover, the stresses $\sigma_h^{\pm}$ are computed as in~\eqref{eq:stressapprox}. We also observe that the set of equations provides a total of $N=n_\delta+n_1+n_2+3$ equations, which are necessary to solve for the $N$ degrees of freedom. The assembled stiffness matrix is shown in Figure~\ref{fig:VHCMmatrix}. Finally, we note that one would recover the exact same system presented in~\cite{diehl2022coupling} if $x_a=\underline{x}_{n_1}=x_0$ and $x_b=\underline{x}_{n_1+1}=x_{n_\delta}$.

\section{Numerical Examples}
\label{sec:numericalexamples}

All numerical examples will be computed with $\ell=3$, $a=1$, and $b=2$, using the configuration shown in Figure~\ref{Fig:peridynamicsdomains}. We will consider the model problem~\eqref{eq:1dlinearelasticity} with the homogeneous Dirichlet boundary condition~\eqref{eq:Dirichlet} at $x=0$ and the Neumann boundary condition~\eqref{eq:Neumann} at $x=\ell$. 

\subsection{Choice of the interpolation operators}

The objective in this section is to highlight the fact that the choice of the interpolation operators may introduce additional errors in the coupling approaches depending on the solution to the problem. It was shown in~\cite{diehl2022coupling} that the three coupling methods have an order of precision of three, meaning that they provide the same solution as that to the local model if the exact solution is a polynomial function of degree three. In the case of discretization with non-matching grids, one should thus use an interpolating polynomial of degree three to maintain the precision of the methods. 

We set $E=1$ and will consider the linear, quadratic, cubic, and quartic manufactured solutions: 
\begin{align}
\label{eq:manufactured:linear:n}
&\underline{u}(x) = \frac{1}{3}x, \\
\label{eq:manufactured:quadratic:n}
&\underline{u}(x) = \frac{1}{9}x^2, \\
\label{eq:manufactured:cubic:n}
&\underline{u}(x) = \frac{1}{27}x^3, \\
\label{eq:manufactured:quartic:n}
&\underline{u}(x) = \frac{1}{81}x^4,
\end{align}
for which the loading function $f_b$ and Neumann boundary condition data $g$ are determined, respectively, as
\begin{align*}
&f_b(x) = -\underline{u}''(x) = 0, &&
g=\underline{u}'(\ell)= \frac{1}{3}, 
\\
&f_b(x) = -\underline{u}''(x) = - \frac{2}{9}, &&
g=\underline{u}'(\ell)=\frac{2}{9}\ell = \frac{2}{3}, 
\\
&f_b(x) = -\underline{u}''(x) = - \frac{2}{9}x, && 
g=\underline{u}'(\ell)=\frac{3}{27}\ell^2 = 1,
\\
&f_b(x) = -\underline{u}''(x) = - \frac{4}{27}x^2, &&
g=\underline{u}'(\ell) = \frac{4}{81}\ell^3=\frac{4}{3}.
\end{align*}
We note that all solutions attain the maximal value of one at $x=\ell$.
For the discretization of the coupling methods, we set the grid sizes in terms of $\delta$ taking $m=2$ and $h_e/h_\delta=2$, that is, $h_\delta=\delta/m = \delta/2$ and $h_e=2h_\delta =\delta$. It therefore implies here that the grids match at the interface points $x=a$ and $x=b$. It also means that VHCM is equivalent in this case to the formulation presented in~\cite{diehl2022coupling}. For illustration purposes, we now provide the entries associated with the coupling constraints applied in $\Gamma_a$ for MDCM and MSCM, namely the top grey blocks in Figures~\ref{fig:MDCMmatrix} and~\ref{fig:MSCMmatrix}. 
We show the case of the quadratic interpolant in Figure~\ref{fig:constraintsI2} and of the cubic interpolant in Figure~\ref{fig:constraintsI3}.

\begin{figure*}[tb]
\centering
\scriptsize
\begin{tikzpicture}
\matrix (m1)[
    matrix of math nodes,
    nodes in empty cells,
    nodes={text width={width(999999)}, 
    align=right},
    right delimiter=\rbrack,left delimiter=\lbrack
  ] at (8,0) {
0 & 1 & 0 & 0 & -1 & 0 & 0 & 0 & 0\\
0 & -1/8 & 6/8 & 3/8 & 0 & -1 & 0 & 0 & 0\\
0 & 0 & 0 & 1 & 0 & 0 & -1 & 0 & 0\\
} ;
\begin{pgfonlayer}{myback}
\fhighlight[cadetgrey!30]{m1-1-2}{m1-3-7}
\end{pgfonlayer}
\matrix (m2)[
    matrix of math nodes,
    nodes in empty cells,
    nodes={text width={width(999999)}, 
    align=right},
    right delimiter=\rbrack,left delimiter=\lbrack
  ] at (8,-1.5) {
0 & -3/2 & 0 & 3/2 & 11 & -18 & 9 & -2 & 0\\ 
0 & 0 & -3 & 3 & 0 & 11 & -18 & 9 & -2 \\
0 & 0 & 0 & 1 & 0 & 0 & -1 & 0 & 0 \\
} ;
\begin{pgfonlayer}{myback}
\fhighlight[cadetgrey!30]{m2-1-2}{m2-3-9}
\end{pgfonlayer}
\end{tikzpicture}
\caption{Coupling equations using the quadratic interpolation operator $I_2$ appearing in the stiffness matrices associated with MDCM (top) and MSCM (bottom). Here, $m=2$, $h_e/h_\delta=2$.}
\label{fig:constraintsI2}
\end{figure*}

\begin{figure*}[tb]
\centering
\scriptsize
\begin{tikzpicture}
\matrix (m1)[
    matrix of math nodes,
    nodes in empty cells,
    nodes={text width={width(999999)}, 
    align=right},
    right delimiter=\rbrack,left delimiter=\lbrack
  ] at (8,0) {
0 & 1 & 0 & 0 & -1 & 0 & 0 & 0 & 0\\
1/16 & -5/16 & 15/16 & 5/16 & 0 & -1 & 0 & 0 & 0\\
0 & 0 & 0 & 1 & 0 & 0 & -1 & 0 & 0\\
} ;
\begin{pgfonlayer}{myback}
\fhighlight[cadetgrey!30]{m1-1-1}{m1-3-7}
\end{pgfonlayer}
\matrix (m2)[
    matrix of math nodes,
    nodes in empty cells,
    nodes={text width={width(999999)}, 
    align=right},
    right delimiter=\rbrack,left delimiter=\lbrack
  ] at (8,-1.5) {
1/2 & -3 & 3/2 & 1 & 11 & -18 & 9 & -2 & 0\\ 
1/8 & -3/8 & -21/8 & 23/8 & 0 & 11 & -18 & 9 & -2 \\
0 & 0 & 0 & 1 & 0 & 0 & -1 & 0 & 0 \\
} ;
\begin{pgfonlayer}{myback}
\fhighlight[cadetgrey!30]{m2-1-1}{m2-3-9}
\end{pgfonlayer}
\end{tikzpicture}
\caption{Coupling equations using the cubic interpolation operator $I_3$ appearing in the stiffness matrices associated with MDCM (top) and MSCM (bottom). Here, $m=2$, $h_e/h_\delta=2$.}
\label{fig:constraintsI3}
\end{figure*}

\begin{figure}[tb]
\centering
\includegraphics[width=0.9\linewidth]{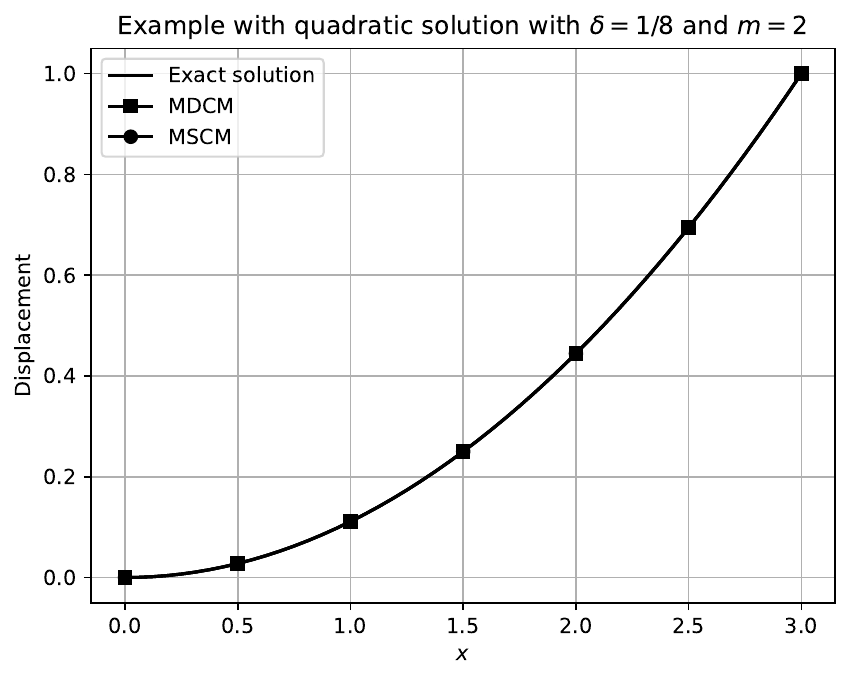}
\caption{Quadratic displacement solution~\eqref{eq:manufactured:quadratic:n} using quadratic interpolation. The coupled solutions obtained by MDCM and MSCM are computed here with horizon $\delta={1}/{8}$, $m=2$, and grid sizes $h_\delta={\delta}/{m}=1/16$ and $h_e=\delta=1/8$.}  
\label{fig:quadratic:solutions}
\end{figure}

\begin{figure}[tb]
\centering
\includegraphics[width=0.9\linewidth]{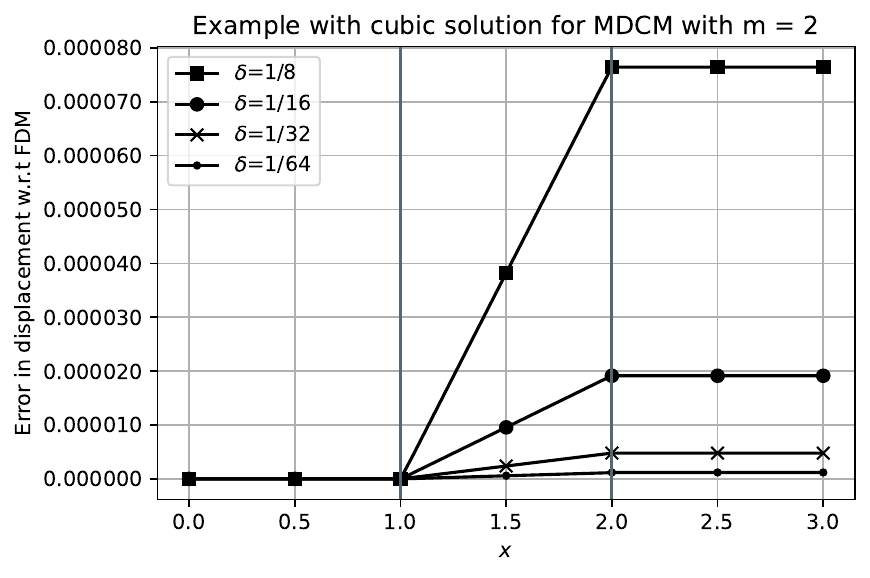}
\includegraphics[width=0.9\linewidth]{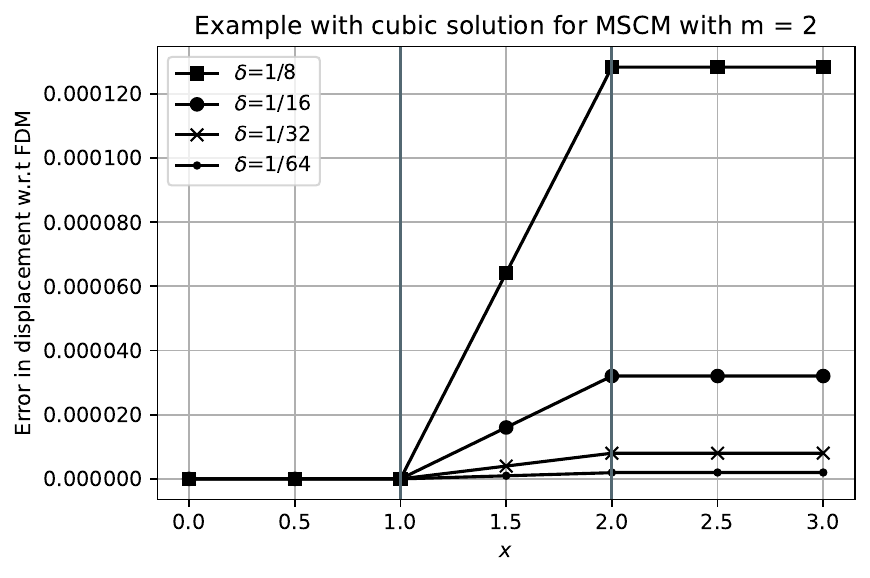}
\caption{Pointwise errors for MDCM (top) and MSCM (bottom) using quadratic interpolation for the problem with a cubic solution~\eqref{eq:manufactured:cubic:n}  and various values of $\delta$. Due to the quadratic interpolation, the degree of precision of the coupling methods is reduced to two.}
\label{fig:quadratic:interpolation:cubic:n}
\end{figure}

\begin{figure}[tb]
\centering
\includegraphics[width=0.9\linewidth]{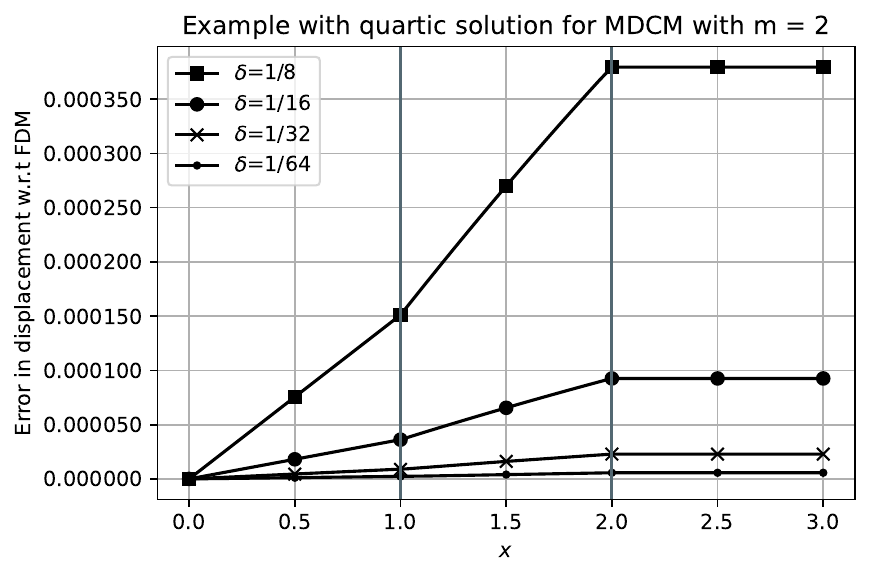}
\includegraphics[width=0.9\linewidth]{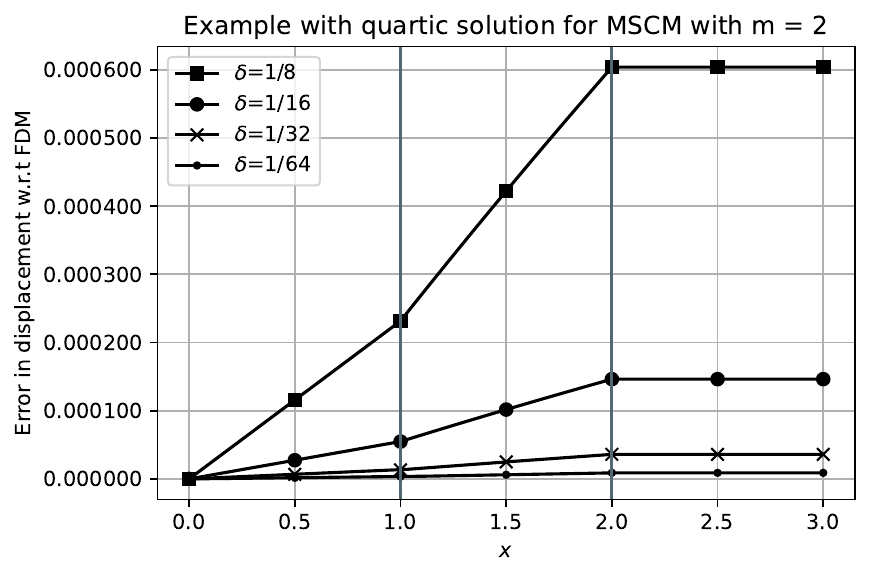}
\includegraphics[width=0.9\linewidth]{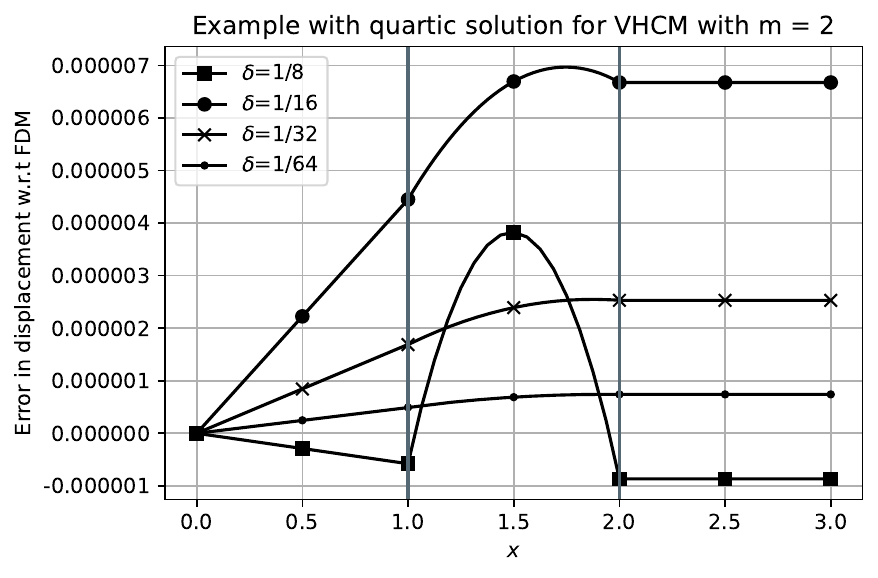}
\caption{Pointwise errors for MDCM (top) and MSCM (middle) using quadratic interpolation for the problem with a quartic solution~\eqref{eq:manufactured:quartic:n} and various values of $\delta$. The degree of precision for MDCM and MSCM is two due to interpolation errors. For comparison, errors in VHCM (bottom) are much smaller than those in MDCM and MSCM since interpolation is not needed here and the degree of precision is therefore three.}
\label{fig:quadratic:interpolation:quartic:n}
\end{figure}

We first consider the case of the quadratic interpolant $I_2(x)$ for MDCM and MSCM. Figure~\ref{fig:quadratic:solutions} confirms that MDCM and MSCM computed here with $\delta=1/8$ exactly recover the quadratic solution, as expected, since interpolation does not introduce any additional errors in the coupling approaches. On the other hand, if the solution to the problem is given by the cubic function~\eqref{eq:manufactured:cubic:n}, one should expect to have errors when using quadratic interpolation. We show in Figure~\ref{fig:quadratic:interpolation:cubic:n} the pointwise errors in the coupled solutions with respect to the approximation to~\eqref{eq:1dlinearelasticity} obtained by the finite difference method (FDM) using $h_e$ everywhere. Due to the quadratic interpolation, the degree of precision of MDCM and MSCM is now reduced to two. We repeat the same experiment in the case of the quartic solution~\eqref{eq:manufactured:quartic:n} and plot the pointwise errors in Figure~\ref{fig:quadratic:interpolation:quartic:n}. In this case, the numerical errors result from a combination of modeling, discretization, and interpolation errors. However, the interpolation errors are dominant. This can be inferred by looking at the coupled solution using VHCM. Indeed, as observed in Figure~\ref{fig:quadratic:interpolation:quartic:n} (bottom), the errors in VHCM are much smaller than those in MDCM and MSCM since the method is void of interpolation errors. Moreover, the errors are only due in this case to modeling and discretization errors. 

\begin{figure}[tb]
\centering
\includegraphics[width=0.9\linewidth]{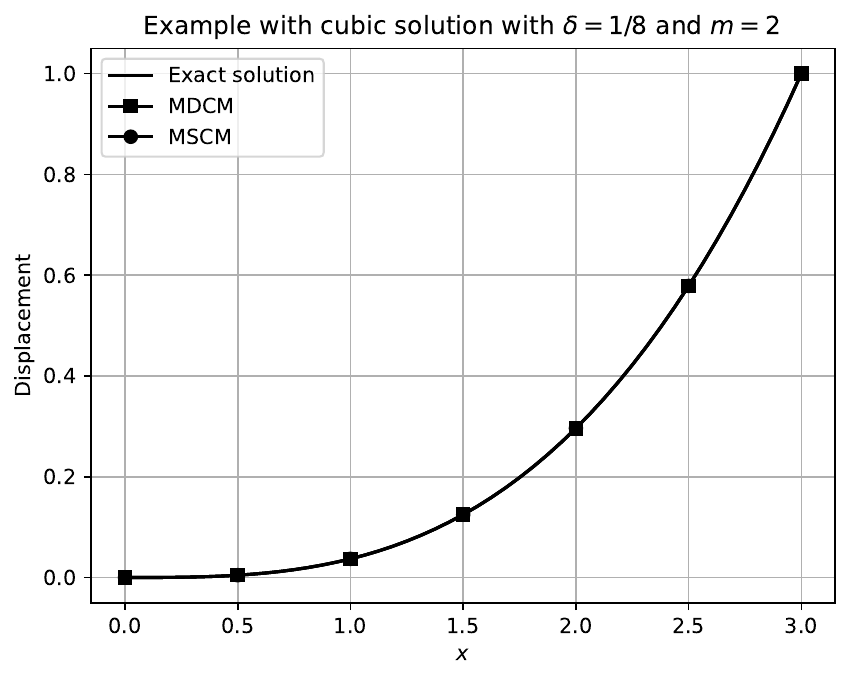}
\caption{Cubic displacement~\eqref{eq:manufactured:cubic:n} for the problem with mixed boundary conditions using cubic interpolation. The coupled solutions obtained by MDCM and VHCM are computed here with horizon $\delta={1}/{8}$, $m=2$, and grid sizes $h_\delta={\delta}/{m}=1/16$ and $h_e=\delta=1/8$.}  
\label{fig:cubic:solutions}
\end{figure}

We now consider the cubic interpolation operator~$I_3$ that allows one to recover the precision of degree three for the three coupling methods. We illustrate in  Figure~\ref{fig:cubic:solutions} that one does indeed recover the exact cubic solution when using MDCM and VHCM, which are shown here for $\delta=1/8$. 

Finally, we consider the case of the quartic solution~\eqref{eq:manufactured:quartic:n}. 
We compile in Table~\ref{tab:cubic:compare:interpolation:n} the maximal values of the pointwise error in the displacement with respect to the FDM solution, for VHCM as well as for MDCM using quadratic interpolation and cubic interpolation. We observe that the errors in the MDCM solution with cubic interpolation are now similar to those obtained with VHCM, in particular for small values of $\delta$, and that the errors in the MDCM solutions using $I_3$ are one order of magnitude smaller that those in the MDCM solutions obtained with $I_2$. We shall use from now on the cubic interpolation operator~$I_3$ only. 

\begin{table}[tb]
\centering
\begin{tabular}{cccc}
\toprule
$\delta$  & VHCM & MDCM with $I_2$ & MDCM with $I_3$ \\
\midrule
1/8  & \num{0.0000038} & \num{0.0003796} & \num{0.0000446} \\
1/16 & \num{0.0000070} & \num{0.0000926} & \num{0.0000123} \\
1/32 & \num{0.0000025} & \num{0.0000229} & \num{0.0000032} \\
1/64 & \num{0.0000007} & \num{0.0000057} & \num{0.0000008} \\
\bottomrule
\end{tabular}
\caption{Maximal values of the error in the displacement for the VHCM and MDCM solutions with respect to the FDM solution in the case of the problem with a quartic solution. The errors in MDCM using $I_3$ are one order of magnitude smaller than those in MDCM obtained using $I_2$.}
\label{tab:cubic:compare:interpolation:n}
\end{table}



\subsection{Examples with a different grid size ratio}
\label{sec:ratio}

\begin{figure}[tb]
\centering
\begin{tikzpicture}[scale=1.4]
\draw (0,0) -- (5.5,0);
\draw (0,0.45) -- (4,0.45);
\draw (3.0,0.9) -- (5.5,0.9);
\draw (3.0,0.0) -- (3.0,1.4);
\draw (4.0,0.0) -- (4.0,1.4);
\draw[arrows=<->, >=stealth] (3.0,1.25) -- (4.0,1.25);
\foreach \i in {0,1}
{\draw (1.5+2.5*\i-0.1,-0.1) -- (1.5+2.5*\i+0.1,0.1);
\draw (1.5+2.5*\i+0.1,-0.1) -- (1.5+2.5*\i-0.1,0.1);}
\foreach \i in {0,1,2}
{\draw (3.0+0.5*\i,-0.1) -- (3.0+0.5*\i,0.1);}
\foreach \i in {0,1}
{\draw[arrows=<-, >=stealth] (3.0+0.5*\i,0.45) -- (3.0+0.5*\i,0.9);}
    
\node[below] at (1.5,-0.5)  {$\underline{x}_{n_1-1}$};
\node[below] at (4.0,-0.5)  {$\underline{x}_{n_1}$};
\node[below] at (3.0,-0.2)  {$x_0$};
\node[below] at (3.5,-0.2)  {$x_1$};
\node[below] at (4.0,-0.2)  {$x_m$};
\node[above] at (1.0,0.6) {$\Omega_1$};
\node[above] at (3.5,1.4) {$\Gamma_a$};
\node[above] at (4.75,1.05) {$\Omega_\delta$};
    
\foreach \i in {0,1}
{\node[circle,color=black,fill=black,inner sep=0pt,minimum size=5pt,label=below:{}] at (1.5+2.5*\i,0.45) {};}

\foreach \i in {0,...,4}
{\node[circle,draw=black,fill=white,inner sep=0pt,minimum size=5pt,label=below:{}] at (3.0+0.5*\i,0.90) {};}
\end{tikzpicture}
\caption{Configuration of the coupling region $\Gamma_a$ for the numerical example of Section~\ref{sec:ratio}. Here, $m=2$, $h_e/h_\delta=5$, and $h_e=a/n=1/n$.}
\label{fig:interface-Gammaa-ratio5}
\end{figure}
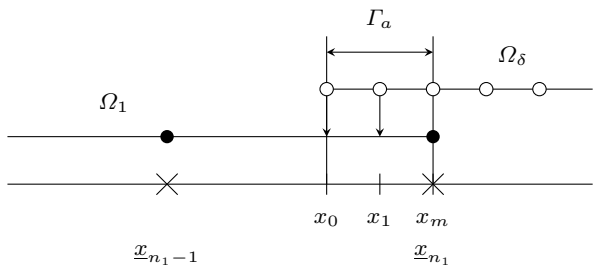

In the previous examples, we assumed that the nodal spacing $h_e$ used for the classical linear elasticity model was twice the grid size $h_\delta$ employed in the peridynamic region. We consider here a larger ratio between the two grid sizes, that is $h_e/h_\delta=5$. We chose a ratio of five in the nodal spacing so that the solution is very accurate in the non-local region without being too fine to be computationally too expensive. More specifically, we keep $m=2$ and choose $h_e={1}/{n}$, where $n$ represents the number of elements in $\Omega_1$ and $\Omega_2$, so that $h_\delta=1/(5n)$ and $\delta=m/(5n)=2/(5n)$, with $n=5,6,7,8$. Moreover, we assume that the end points $\underline{x}_{n_1}$ and $x_m$ still coincide at $x=a$ and that the end points $\underline{x}_{n_1+1}$ and $x_{m+n_\delta}$ also coincide at $x=b$. The corresponding configuration of the coupling region~$\Gamma_a$ is shown in Figure~\ref{fig:interface-Gammaa-ratio5}.

\begin{figure}[tb]
\includegraphics[width=0.95\linewidth]{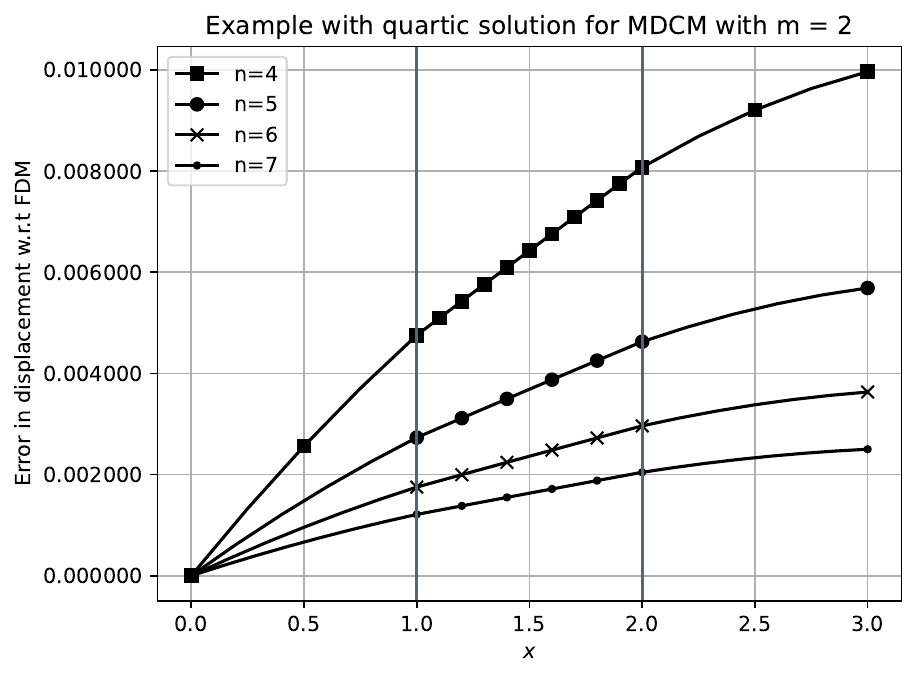}
\caption{Pointwise error in the MDCM solution using cubic interpolation with respect to the FDM solution in the case of the quartic manufactured solution~\eqref{eq:manufactured:quartic:n}. Here, $n$ denotes the number of elements in each subdomain $\Omega_1$ and $\Omega_2$, that is $h_e=1/n$, and $h_e/h_\delta=5$.}
\label{fig:variation:h:match:interface}
\end{figure}

Figure~\ref{fig:variation:h:match:interface} shows the pointwise error in the approximation of the quartic solution by MDCM using cubic interpolation with respect to the FDM solution computed with $h_\delta$ everywhere in $\Omega$. As expected, the errors are larger in magnitude than those obtained in Figure~\ref{fig:quadratic:interpolation:quartic:n} since the reference solution is computed here with respect to the FDM solution using a smaller grid size. However, we have checked that we still recover the exact solution for the manufactured solutions up to degree three. Moreover, we observe that the errors decrease as the grid size decreases.

\subsection{Example with misaligned grids at interface points}
\label{fig:epsilon}

\begin{figure}[tb]
\centering
\begin{tikzpicture}[scale=1.4]
\draw (1.0,0.0) -- (6.0,0.0);
\draw (1,0.45) -- (4,0.45);
\draw (3.25,0.9) -- (6.0,0.9);
\draw (3.25,0.0) -- (3.25,1.4);
\draw (4.00,0.0) -- (4.00,1.4);
\draw (4.25,0.0) -- (4.25,1.4);
\draw[arrows=<->, >=stealth] (3.25,1.25) -- (4.00,1.25);
\draw[arrows=<->, >=stealth] (4.00,1.25) -- (4.25,1.25);
\foreach \i in {0,1}
{\draw (1.5+2.5*\i-0.1,-0.1) -- (1.5+2.5*\i+0.1,0.1);
\draw (1.5+2.5*\i+0.1,-0.1) -- (1.5+2.5*\i-0.1,0.1);}
\foreach \i in {0,1,2}
{\draw (3.25+0.5*\i,-0.1) -- (3.25+0.5*\i,0.1);}
\foreach \i in {0,1}
{\draw[arrows=<-, >=stealth] (3.25+0.5*\i,0.45) -- (3.25+0.5*\i,0.9);}
\draw[arrows=->, >=stealth] (4.0,0.45) -- (4.0,0.9);
    
\node[below] at (1.5,-0.5)  {$\underline{x}_{n_1-1}$};
\node[below] at (4.0,-0.5)  {$\underline{x}_{n_1}$};
\node[below] at (4.0,-0.1)  {$a$};
\node[below] at (3.25,-0.3)  {$x_0$};
\node[below] at (3.75,-0.3)  {$x_1$};
\node[below] at (4.25,-0.3)  {$x_m$};
\node[above] at (2.25,0.6) {$\Omega_1$};
\node[above] at (3.625,1.4) {$\Gamma_a$};
\node[above] at (4.125,1.4) {$\varepsilon$};
\node[above] at (5.00,1.05) {$\Omega_\delta$};
    
\foreach \i in {0,1}
{\node[circle,color=black,fill=black,inner sep=0pt,minimum size=5pt,label=below:{}] at (1.5+2.5*\i,0.45) {};}

\foreach \i in {0,...,5}
{\node[circle,draw=black,fill=white,inner sep=0pt,minimum size=5pt,label=below:{}] at (3.25+0.5*\i,0.90) {};}
\end{tikzpicture}
\caption{Configuration of the coupling region $\Gamma_a$ for the numerical example of Section~\ref{fig:epsilon}. Here, $m=2$, $h_e/h_\delta=5$, $h_e=a/n=1/n$, and $\varepsilon=h_\delta/2$.}
\label{fig:interface-Gammaa-epsilon}
\end{figure}
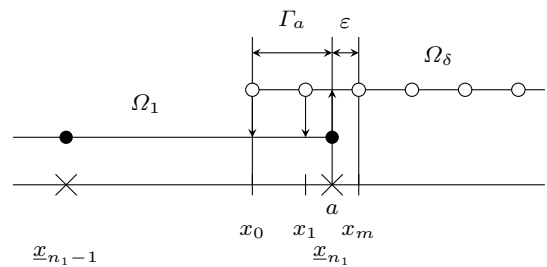

We now study the configuration where the two grids are misaligned at the interface points $x=a$ and $x=b$. We introduce an offset of $\varepsilon$ such that $x_{m}-a = \varepsilon$ and $b - x_{m+n_\delta} = \varepsilon$. The offset is chosen as $\varepsilon = h_\delta/2$. Moreover, we set $m=2$, $h_e/h_\delta=5$, and $h_e=a/n=1/n$, $n=4,\ldots,7$. The set-up is shown in Figure~\ref{fig:interface-Gammaa-epsilon} for the interface region $\Gamma_a$. We note that this configuration is slightly different from that of VHCM shown in Figure~\ref{Fig:discretizationVHCM} in the sense that $x=a$ now coincides with $\underline{x}_{n_1}$ and not $x_m$ (or $x_0$ in the case of VHCM). The motivation here is to ensure that the displacements that need to be interpolated always lie within the points used to determine the interpolation operator~$I_p$. 

\begin{figure}[tb]
\centering
\includegraphics[width=\linewidth]{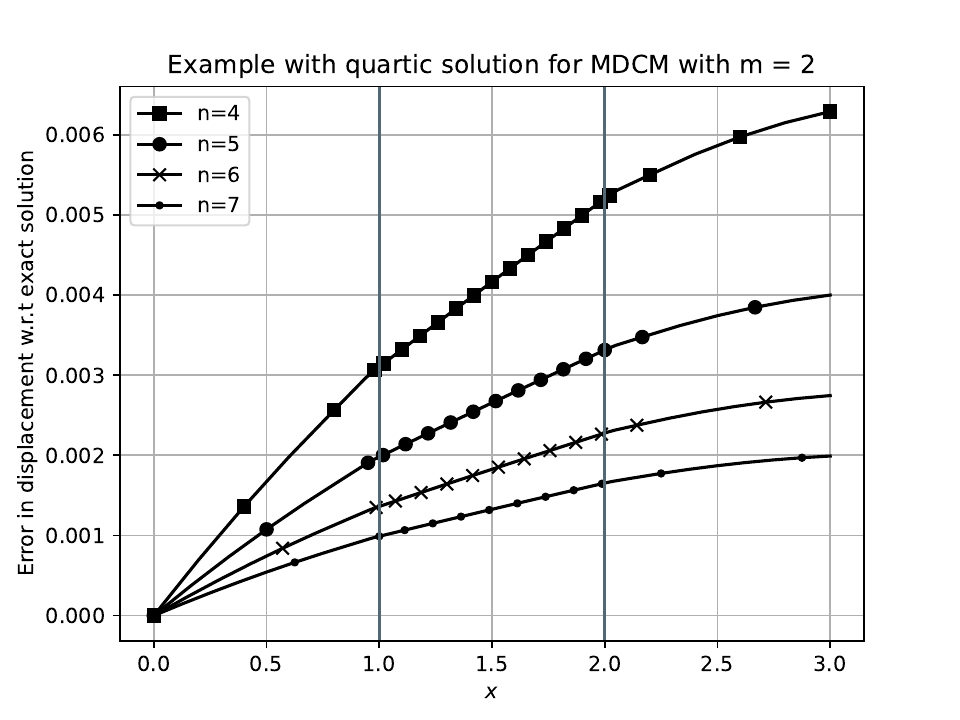}
\caption{Pointwise error in the MDCM approximation using cubic interpolation with respect to the exact solution~\eqref{eq:manufactured:quartic:n}. The nodes at the interfaces $x=a$ and $x=b$ are not aligned. Here, $n$ denotes the number of elements in each subdomain $\Omega_1$ and $\Omega_2$, $m=2$, and $h_e/h_\delta= 5$.}
\label{fig:variation:h:no:match:interface}
\end{figure}

Figure~\ref{fig:variation:h:no:match:interface} shows the pointwise error with respect to the exact solution to the classical linear elasticity model (rather than the FDM solution). As before, we observe that the error decreases as the grid is refined. Moreover, the errors are of the same order as those obtained in Figure~\ref{fig:variation:h:match:interface} showing that misaligned points at the interface do not affect the quality of the solutions. 

\begin{figure}[tb]
\centering
\includegraphics[width=0.9\linewidth]{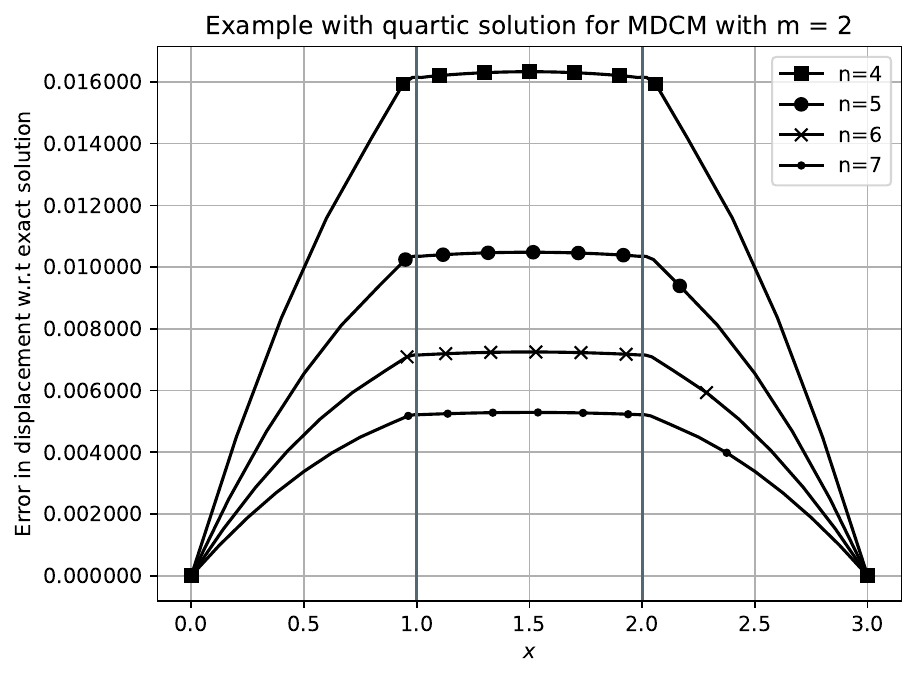}
\caption{Pointwise error in the MDCM approximation using cubic interpolation with respect to the exact  solution~\eqref{eq:uexactDirichletquartic} in the case of homogeneous Dirichlet boundary conditions at both ends of the bar. The nodes at the interfaces $x=a$ and $x=b$ are not aligned. Here, $n$ denotes the number of elements in each subdomain $\Omega_1$ and $\Omega_2$, $m=2$, and $h_e/h_\delta= 5$.}
\label{fig:variation:h:no:match:interface:d}
\end{figure}

We repeat the same experiment in the case of homogeneous Dirichlet boundary conditions at both ends of the bar, i.e.\ at $x=0$ and $x=\ell$. The exact quartic solution in that case reads:
\begin{equation}
\label{eq:uexactDirichletquartic}
\underline{u}(x) = \frac{16}{81} x^2 (3-x)^2.
\end{equation}
We note that the maximal value of $\underline{u}(x)$ reaches one at $x=3/2$.
The results are shown in Figure~\ref{fig:variation:h:no:match:interface:d}. We observe that the errors are larger in the central region of the domain compared with the case with a Neumann condition at $x=3$. This is due to the fact that we impose using a Dirichlet condition a stronger constraint on the boundary, which in turn generates higher errors inside the domain.

\subsection{Example with varying modulus of elasticity}

\begin{figure}
\centering
\includegraphics[width=\linewidth]{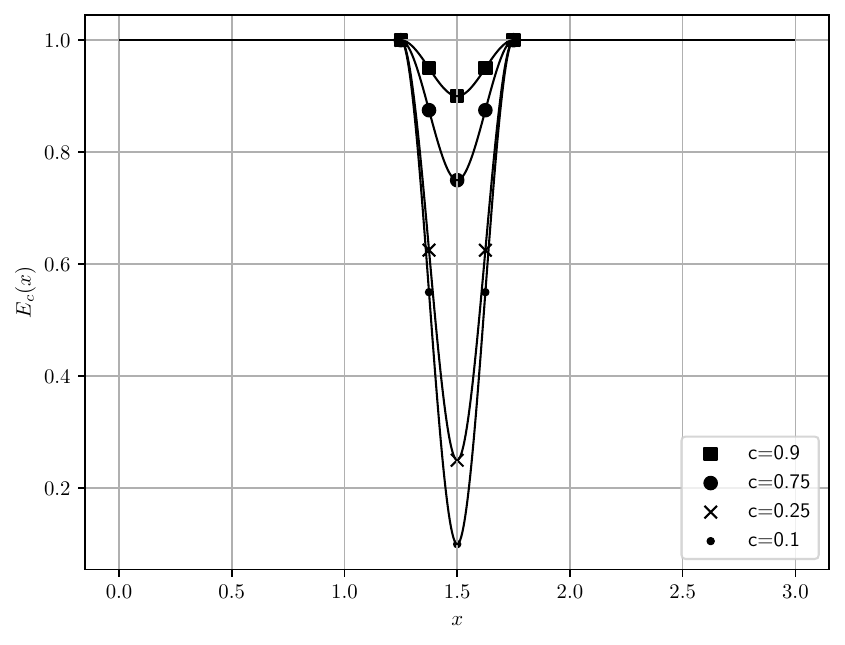}
\caption{Examples of varying $E(x)$ using cubic spline interpolation for $c=0.1$, $0.75$, $0.25$, and $0.9$. The markers represent the data points used for the spline interpolation.}
\label{fig:damage:spline}
\end{figure}

We now study the effect of a smoothly varying modulus of elasticity $E=E(x)$ in order to mimic the presence of a local defect or damage in the bar, justifying the use of a coupling approach. We set $E=1$ for all $x$ within the regions $(0,1.25)$ and $(1.75,2)$ and suppose that $E$ is given by piecewise cubic spline functions in the interval $(1.25,1.75) \subset \Omega_\delta$. We construct the cubic spline functions such that they interpolate the set of data points $(x_i,y_i) \in \{(1.25,1), ((1.25+1.5)/2,(1+c)/2), (1.5,c), ((1.75+1.5)/2,(1+c)/2), (1.75,1)\}$, where $c$ is a scalar between zero and one. We show the resulting $E(x)$ in Figure~\ref{fig:damage:spline} for $c=0.1$, $0.25$, $0.75$, and $1$. The lower the value of $c$, the softer the material becomes at the center of the bar. We compute the solutions using the MDCM coupling approach in the case where the nodes are aligned at the interfaces and the problems with mixed boundary conditions. 
 
We report in Table~\ref{tab:error:damage:n} the maximal pointwise error in the MDCM approximation with respect to the finite difference solution computed with the same grid as that used for the coupled solution, for several values of the horizon $\delta$ and $m=2$. Moreover, we consider the linear, quadratic, cubic, and quartic manufactured solutions. We observe in all cases that the error is divided by about four each time the horizon is halved, at least for small enough values of $\delta$. Moreover, the error systematically increases as the value of $c$ decreases as expected. On the other hand, the error first increases and then decreases when increasing the degree of the exact solutions. 

\begin{table}[tb]
\centering
\sisetup{round-mode=places,round-precision=7,group-separator={}}
\begin{tabular}{ccccc}
$\delta$  & Linear & Quadratic & Cubic & Quartic  \\
\midrule
\multicolumn{5}{c}{c=0.9}\\
\midrule
{1}/{8}  & \num{0.0004743879} & \num{0.0008048903} & \num{0.0008550996} & \num{0.0004092119} \\
{1}/{16} & \num{0.0001200195} & \num{0.0002108269} & \num{0.0002270992} & \num{0.0001320202} \\
{1}/{32} & \num{0.0000304043} & \num{0.0000541946} & \num{0.0000587087} & \num{0.0000366584} \\
{1}/{64} & \num{0.0000076385} & \num{0.0000137248} & \num{0.0000149136} & \num{0.0000096157} \\
\midrule
\multicolumn{5}{c}{c=0.75}\\
\midrule
{1}/{8}  & \num{0.0015207368145301325} & \num{0.0023661730} & \num{0.0024188138} & \num{0.0017200164} \\
{1}/{16} & \num{0.0003964320417785272} & \num{0.0006276174} & \num{0.0006466189} & \num{0.0004849140} \\
{1}/{32} & \num{0.00010113320183424701} & \num{0.0001616099} & \num{0.0001672027} & \num{0.0001281181} \\
{1}/{64} & \num{0.0000254589} & \num{0.0000409240} & \num{0.0001672027} & \num{0.0000328701} \\
\midrule
\multicolumn{5}{c}{c=0.5}\\
\midrule
{1}/{8} & \num{0.0052547932} & 
\num{0.0070306856} & \num{0.0066316345} & \num{0.0050171533} \\
{1}/{16} & \num{0.0014670512} & \num{0.0019497268} & \num{0.0018311628} & \num{0.0014068558} \\
{1}/{32} & \num{0.0003805258} & \num{0.0005065490} & \num{0.0004761879} & \num{0.0003686359} \\
{1}/{64} & \num{0.0000962315} & \num{0.0001284305} & \num{0.0001209166} & \num{0.0000940043} \\
\midrule
\multicolumn{5}{c}{c=0.25}\\
\midrule
{1}/{8}  & \num{0.0189032644} & 
\num{0.0217855300} & \num{0.0185628219} & \num{0.0135670937} \\
{1}/{16} & \num{0.0061267099} & \num{0.0069132368} & \num{0.0057917737} & \num{0.0042111557} \\
{1}/{32} & \num{0.0016585077} & \num{0.0018640318} & \num{0.0015566880} & \num{0.0011320452} \\
{1}/{64} & \num{0.0004239457} & \num{0.0004765221} & \num{0.0003979537} & \num{0.0002897285} \\
\midrule
\multicolumn{5}{c}{c=0.1}\\
\midrule
{1}/{8}  & \num{0.0580608552} & 
\num{0.0618444766} & \num{0.0493397823} & \num{0.0346054167} \\
{1}/{16} & \num{0.0258549055} & \num{0.0269246387} & \num{0.0210321509} & \num{0.0145342177} \\
{1}/{32} & \num{0.0078532931} & \num{0.0081374090} & \num{0.0063260344} & \num{0.0043565707} \\
{1}/{64} & \num{0.0020744831} & \num{0.0021474957} & \num{0.0016679495} & \num{0.0011481485} \\
\end{tabular}
\caption{Maximal error with respect to the FDM solution in the case of various profiles of the modulus of elasticity $E(x)$ using MDCM and cubic interpolation.}
\label{tab:error:damage:n}
\end{table}

\subsection{Condition number of the coupled systems}
\label{sec:coupling:condition}

We study in this section the behavior of the condition number of the stiffness matrices resulting from various coupling configurations.
We use the NumPy function \texttt{numpy.linalg.con}\footnote{\url{https://numpy.org/doc/stable/reference/generated/numpy.linalg.cond.html}} to compute the condition number $\text{cond}(M)$ of the resulting matrix~$M$. The condition number in NumPy is computed as $\text{cond}(M)= \|M\|_{2} \|M^{-1}\|_{2}$, see \emph{e.g.}~\cite{lay2016linear}. All results below are calculated using the problems with mixed boundary conditions at the extremities of the bar.

\paragraph{Non-matching grids with aligned interfaces:}

\begin{figure}[tb]
\centering
\includegraphics[width=0.9\linewidth]{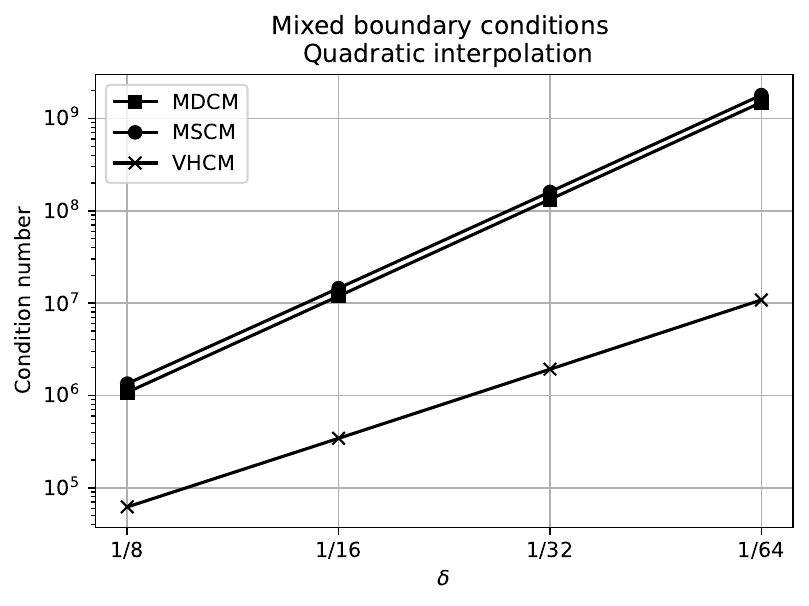}
\caption{Condition number of the stiffness matrix with respect to the horizon $\delta$ for non-matching grids and aligned interfaces.}
\label{fig:condition:number}
\end{figure}

We consider the following discretization $h_e/h_\delta = 5$, $m=2$, with aligned interfaces.
We show in Figure~\ref{fig:condition:number} the evolution of the condition number with respect to the horizon $\delta$ for MDCM, MSCM, and VHCM using quadratic interpolation. The condition number for the three coupling approaches increases with smaller horizons. While it is similar for MDCM and MSCM, we observe that the condition number for VHCM is between one and two orders of magnitude lower than that for MDCM and MSCM.

\paragraph{Non-matching grids and misaligned interface in the case of MDCM:}

\begin{figure}[tb]
\centering
\includegraphics[width=0.9\linewidth]{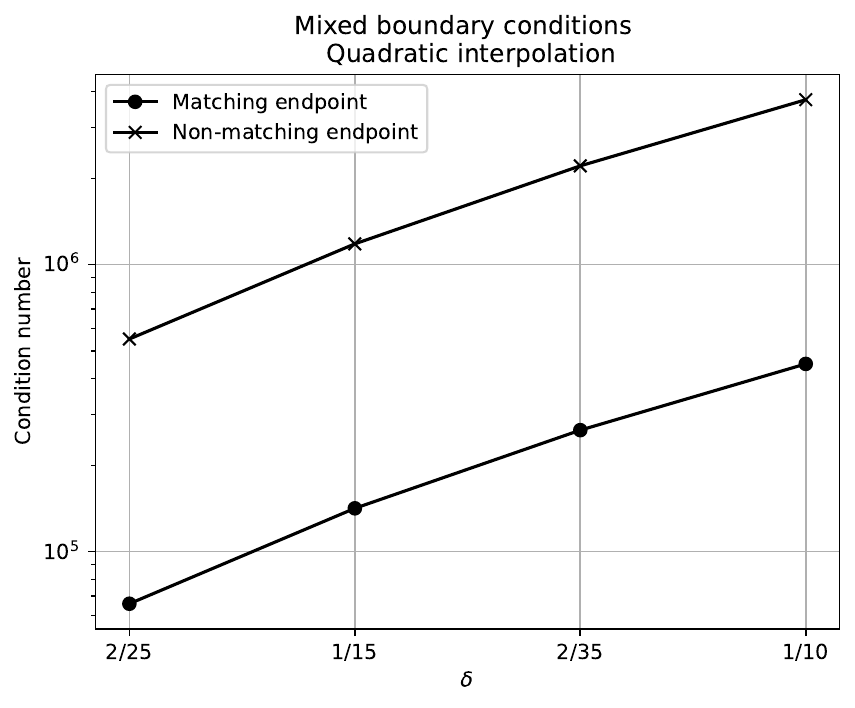}
\caption{Condition number of the stiffness matrix for MDCM using quadratic interpolation with respect to the horizon $\delta$ with non-matching grids and misaligned interfaces.}
\label{fig:condition:number:quadratic}
\end{figure}

\begin{figure}[tb]
\centering
\includegraphics[width=0.9\linewidth]{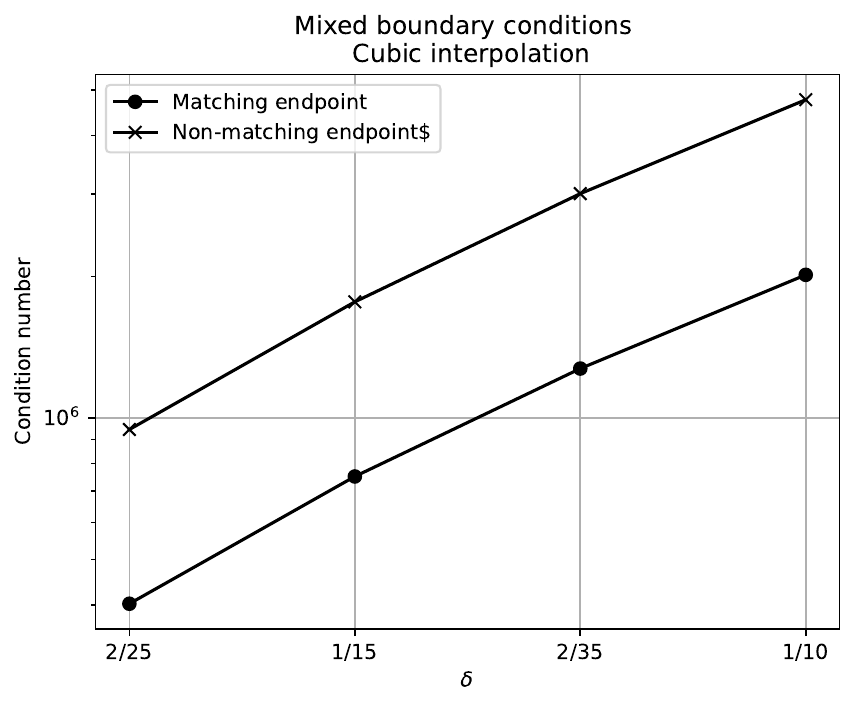}
\caption{Condition number of the stiffness matrix for MDCM using cubic interpolation with respect to the horizon $\delta$ with non-matching grids and misaligned interfaces.}
\label{fig:condition:number:cubic}
\end{figure}

We plot in Figures~\ref{fig:condition:number:quadratic} and~\ref{fig:condition:number:cubic} the condition number for MDCM with respect to the horizon~$\delta$ using quadratic and cubic interpolation, respectively, and $m=2$.
On one hand, we observe that the type of interpolation (quadratic or cubic) has little effect on the values of the condition number. On the other hand, the condition number is slightly higher when the endpoints do not match at the interface.

\paragraph{Varying modulus of elasticity:}

\begin{figure}[tb]
\centering
\includegraphics[width=0.9\linewidth]{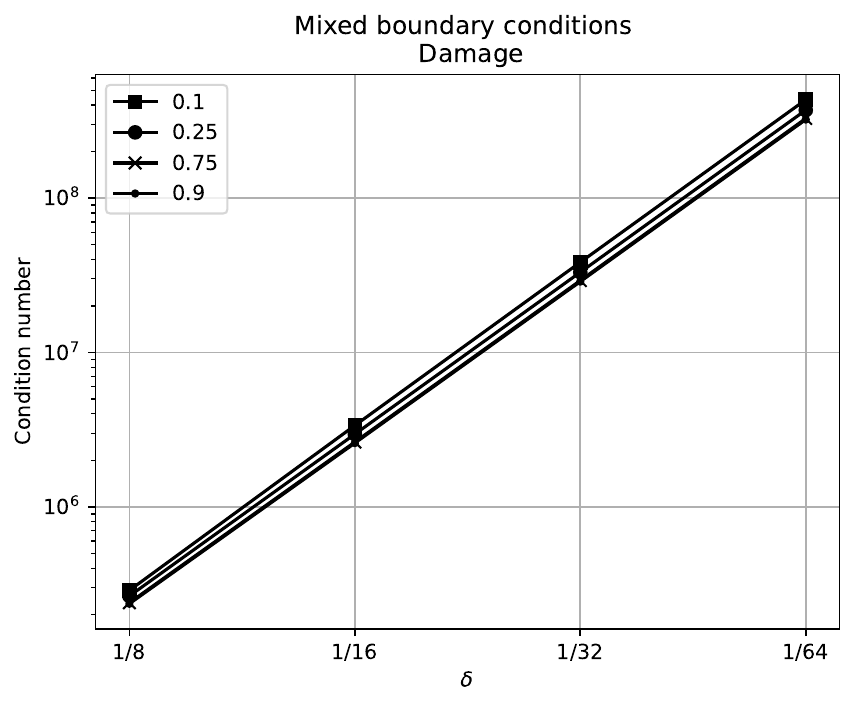}
\caption{Condition number of the stiffness matrix for MDCM with varying $E$ with respect to $\delta$.}
\label{fig:condition:number:damage}
\end{figure}

The condition number of the stiffness matrix obtained using MDCM, as a function of the horizon $\delta$, is shown in Figure~\ref{fig:condition:number:damage} in the cases of varying moduli of elasticity considered in Figure~\ref{fig:damage:spline}. We use here MDCM with matching endpoints at both interfaces and the same grid spacing $h_e=h_\delta$, with $\delta=m h_\delta$ and $m=2$. We observe that the condition numbers are similar for all values of parameter $c$, whether small or large.

\section{Conclusion}
\label{sec:conclusion}

We have presented in this paper three coupling methods, namely the coupling method with matching displacements (MSCM), the coupling method with matching stresses (MSCM), and the variable horizon coupling method (VHCM), for coupling classical linear elasticity and peridynamic models using non-matching grids in one dimension. 

Because one wants in general to use a smaller nodal spacing for the discretization of the peridynamic model than that for the classical linear elasticity model, it is often advantageous to consider non-matching grid points in the overlap regions at the interfaces of the two models. We have suggested here that one could use an interpolation of the linear elasticity displacement fields to be able to match displacements and stresses over the overlap regions, in the MDCM and MSCM approaches, respectively. However, we have shown that the choice of the interpolation scheme was essential to  preserve the accuracy of the coupling methods. For example, we have seen that the degree of the interpolants should be at least $p$ if one did not want to introduce any additional numerical errors when using MDMC and MSMC for solving problems whose exact solution was of degree at most $p$. Moreover, in the case of a quartic solution and cubic interpolation, we have shown on several numerical examples that the error in the MDCM and MSCM solutions decreases as the horizon goes to zero. 
For VHMC, interpolation is not necessary unless the grid points do not match at the interfaces themselves. We have also tested the cubic interpolation scheme with MDCM for non-constant moduli of elasticity. The objective of the study was to assess the well-foundedness of the interpolation scheme for the coupling approaches with non-matching grids when considering materials with local defects. 

Moreover, we have observed that that the condition numbers of the stiffness matrices resulting from the coupled systems for various configurations were not much affected by interpolation when coupling models using non-matching grids. As a general observation, the VHCM always has a smaller condition number than MDCM and MSCM. 

The main objective for future works is to extend the interpolation scheme for coupling models in higher dimensions. It is clear that finding the quadratic or cubic interpolant of the displacement fields in 2D and 3D is not necessarily a trivial task. A possible approach for interpolation would be to define the interpolant in terms of neural networks. We will address this issue in a forthcoming paper. 

\section*{Supplementary materials}
The Python code (using \texttt{NumPy}~\cite{oliphant2006guide,5725236}, \texttt{SymPy}~\cite{10.7717/peerj-cs.103}, and \texttt{Matplotlib}~\cite{Hunter:2007}) used to generate the numerical results without damage are available on GitHub\textsuperscript{\textregistered}\footnote{\url{https://github.com/diehlpk/reusommer21}} and on Zenodo\textsuperscript{\textregistered}~\cite{patrick_diehl_2023_8083971}. The scripts for numerical results including damage are available on GitHub\textsuperscript{\textregistered}\footnote{\url{https://github.com/diehlpk/reusummer22}} and Zenodo\textsuperscript{\textregistered} \cite{patrick_diehl_2023_8061731}.

\begin{acknowledgements}
Part of this research was done by A. Edwards during the Research Experiences for Undergraduates in Physics at Louisiana State University in Summer 2021. Part of this research was done by E. Downing during the Research Experiences for Undergraduates in Physics at Louisiana State University in Summer 2022.
\end{acknowledgements}

\section*{Declarations}

\subsection*{Ethical Approval}
Not applicable.

\subsection*{Competing interests}
The authors have no competing interests.

\subsection*{Authors' contributions}
ED and AE contributed to the Python codes during their research experience for undergraduates at Louisiana State University at PD prior appointment. PD and SP supervised the two students. ED, AE, and PD generated the figures for the paper. PD and SP wrote the manuscript and conceptualized the research.

\subsection*{Funding}
 This material is based upon work supported by the National Science Foundation under award OAC-2150491 with additional support from the Center for Computation \& Technology at Louisiana State University. This work was supported by the U.S. Department of Energy through the Los Alamos National Laboratory. Los Alamos National Laboratory is operated by Triad National Security, LLC, for the National Nuclear Security Administration of U.S. Department of Energy (Contract No.\ 89233218CNA000001).
SP is grateful for the support from the Natural Sciences and Engineering Research Council of Canada (NSERC) Discovery Grant [Grant No.\ RGPIN-2019-7154]. LA-UR-24-32675 (Revision 1)


\bibliographystyle{ieeetr}
\bibliography{references}

\begin{thebibliography}{10}

\bibitem{d2021review}
M.~D’Elia, X.~Li, P.~Seleson, X.~Tian, and Y.~Yu, ``A review of
  local-to-nonlocal coupling methods in nonlocal diffusion and nonlocal
  mechanics,'' {\em Journal of Peridynamics and Nonlocal Modeling}, pp.~1--50,
  2021.

\bibitem{diehl2019review}
P.~Diehl, S.~Prudhomme, and M.~L{\'e}vesque, ``A review of benchmark
  experiments for the validation of peridynamics models,'' {\em Journal of
  Peridynamics and Nonlocal Modeling}, vol.~1, no.~1, pp.~14--35, 2019.

\bibitem{du2016nonlocal}
Q.~Du, ``Nonlocal calculus of variations and well-posedness of peridynamics,''
  in {\em Handbook of Peridynamic Modeling}, pp.~101--124, Chapman and
  Hall/CRC, 2016.

\bibitem{madenci2018state}
E.~Madenci, M.~Dorduncu, A.~Barut, and N.~Phan, ``A state-based peridynamic
  analysis in a finite element framework,'' {\em Engineering Fracture
  Mechanics}, vol.~195, pp.~104--128, 2018.

\bibitem{madenci2018weak}
E.~Madenci, M.~Dorduncu, A.~Barut, and N.~Phan, ``Weak form of peridynamics for
  nonlocal essential and natural boundary conditions,'' {\em Computer Methods
  in Applied Mechanics and Engineering}, vol.~337, pp.~598--631, 2018.

\bibitem{gu2018revisit}
X.~Gu, E.~Madenci, and Q.~Zhang, ``Revisit of non-ordinary state-based
  peridynamics,'' {\em Engineering Fracture Mechanics}, vol.~190, pp.~31--52,
  2018.

\bibitem{Prudhomme-Diehl-2020}
S.~Prudhomme and P.~Diehl, ``On the treatment of boundary conditions for
  bond-based peridynamic models,'' {\em Computer Methods in Applied Mechanics
  and Engineering}, vol.~372, p.~113391, 2020.

\bibitem{you2020asymptotically}
H.~You, X.~Lu, N.~Task, and Y.~Yu, ``An asymptotically compatible approach for
  neumann-type boundary condition on nonlocal problems,'' {\em ESAIM:
  Mathematical Modelling and Numerical Analysis}, vol.~54, no.~4,
  pp.~1373--1413, 2020.

\bibitem{d2021prescription}
M.~D’Elia and Y.~Yu, ``On the prescription of boundary conditions for
  nonlocal poisson’s and peridynamics models,'' in {\em Research in
  Mathematics of Materials Science}, pp.~185--207, Springer, 2022.

\bibitem{d2020physically}
M.~D'Elia, X.~Tian, and Y.~Yu, ``A physically consistent, flexible, and
  efficient strategy to convert local boundary conditions into nonlocal volume
  constraints,'' {\em SIAM Journal on Scientific Computing}, vol.~42, no.~4,
  pp.~A1935--A1949, 2020.

\bibitem{zhao2025enforcing}
J.~Zhao, S.~Jafarzadeh, Z.~Chen, and F.~Bobaru, ``Enforcing local boundary
  conditions in peridynamic models of diffusion with singularities and on
  arbitrary domains,'' {\em Engineering with Computers}, vol.~41, no.~1,
  pp.~247--266, 2025.

\bibitem{diehl2022comparative}
P.~Diehl, R.~Lipton, T.~Wick, and M.~Tyagi, ``A comparative review of
  peridynamics and phase-field models for engineering fracture mechanics,''
  {\em Computational Mechanics}, pp.~1--35, 2022.

\bibitem{zaccariotto2018coupling}
M.~Zaccariotto, T.~Mudric, D.~Tomasi, A.~Shojaei, and U.~Galvanetto,
  ``{Coupling of FEM meshes with Peridynamic grids},'' {\em Computer Methods in
  Applied Mechanics and Engineering}, vol.~330, pp.~471--497, 2018.

\bibitem{zaccariotto2017enhanced}
M.~Zaccariotto, D.~Tomasi, and U.~Galvanetto, ``{An enhanced coupling of PD
  grids to FE meshes},'' {\em Mechanics Research Communications}, vol.~84,
  pp.~125--135, 2017.

\bibitem{diehl2022coupling}
P.~Diehl and S.~Prudhomme, ``Coupling approaches for classical linear
  elasticity and bond-based peridynamic models,'' {\em {Journal of Peridynamics
  and Nonlocal Modeling}}, pp.~1--31, 2022.

\bibitem{Silling-JMPS-2000}
S.~A. Silling, ``Reformulation of elasticity theory for discontinuities and
  long-range forces,'' {\em Journal of the Mechanics and Physics of Solids},
  vol.~48, pp.~175--209, 2000.

\bibitem{DElia-Bochev-2021}
M.~D'Elia and P.~Bochev, ``Formulation, analysis and computation of an
  optimization-based local-to-nonlocal coupling method,'' {\em Results in
  Applied Mathematics}, vol.~9, p.~100129, 2021.

\bibitem{silling2020Couplingstresses}
S.~Silling, ``Local-nonlocal coupling in {Emu/PDMS},'' {\em Sandia Report
  SAND2020-11382}, 2020.

\bibitem{https://doi.org/10.1002/nme.2439}
F.~Bobaru, M.~Yang, L.~F. Alves, S.~A. Silling, E.~Askari, and J.~Xu,
  ``Convergence, adaptive refinement, and scaling in 1d peridynamics,'' {\em
  International Journal for Numerical Methods in Engineering}, vol.~77, no.~6,
  pp.~852--877, 2009.

\bibitem{silling2015variable}
S.~Silling, D.~Littlewood, and P.~Seleson, ``Variable horizon in a peridynamic
  medium,'' {\em Journal of Mechanics of Materials and Structures}, vol.~10,
  no.~5, pp.~591--612, 2015.

\bibitem{NIKPAYAM2019308}
J.~Nikpayam and M.~A. Kouchakzadeh, ``A variable horizon method for coupling
  meshfree peridynamics to {FEM},'' {\em Computer Methods in Applied Mechanics
  and Engineering}, vol.~355, pp.~308--322, 2019.

\bibitem{10.7717/peerj-cs.103}
A.~Meurer, C.~P. Smith, M.~Paprocki, O.~\v{C}ert\'{i}k, S.~B. Kirpichev,
  M.~Rocklin, A.~Kumar, S.~Ivanov, J.~K. Moore, S.~Singh, T.~Rathnayake,
  S.~Vig, B.~E. Granger, R.~P. Muller, F.~Bonazzi, H.~Gupta, S.~Vats,
  F.~Johansson, F.~Pedregosa, M.~J. Curry, A.~R. Terrel, v.~Rou\v{c}ka,
  A.~Saboo, I.~Fernando, S.~Kulal, R.~Cimrman, and A.~Scopatz, ``Sympy:
  symbolic computing in python,'' {\em PeerJ Computer Science}, vol.~3,
  p.~e103, Jan. 2017.

\bibitem{Bilodeau2024}
C.~Bilodeau, P.~Diehl, R.~Flachaire, and S.~Prudhomme, ``High-order integration
  rules for peridynamic modeling in one and two dimensions,'' 2024.
\newblock In prepration.

\bibitem{lay2016linear}
D.~C. Lay, S.~R. Lay, and J.~J. McDonald, {\em Linear algebra and its
  applications}.
\newblock Pearson, 2016.

\bibitem{oliphant2006guide}
T.~E. Oliphant, {\em A guide to NumPy}, vol.~1.
\newblock Trelgol Publishing USA, 2006.

\bibitem{5725236}
S.~{van der Walt}, S.~C. {Colbert}, and G.~{Varoquaux}, ``The {NumPy} {Array}:
  A structure for efficient numerical computation,'' {\em Computing in Science
  \& Engineering}, vol.~13, no.~2, pp.~22--30, 2011.

\bibitem{Hunter:2007}
J.~D. Hunter, ``Matplotlib: A 2d graphics environment,'' {\em Computing in
  Science \& Engineering}, vol.~9, no.~3, pp.~90--95, 2007.

\bibitem{patrick_diehl_2023_8083971}
P.~Diehl and A.~Edwards, ``{Coupling approaches with non-matching grids for
  classical linear elasticity and bond-based peridynamic models in 1D - Part
  I}.'' https://doi.org/10.5281/zenodo.8083971, June 2023.

\bibitem{patrick_diehl_2023_8061731}
P.~Diehl and E.~Downing, ``{Coupling approaches with non-matching grids for
  classical linear elasticity and bond-based peridynamic models in 1D - Part
  II}.'' https://doi.org/10.5281/zenodo.8061731, June 2023.

\end{thebibliography}


\end{document}